\documentclass[lettersize,journal]{IEEEtran}
\usepackage[table,xcdraw]{xcolor}
\definecolor{lightgrey}{HTML}{808080}

\usepackage{amsmath,amsfonts,amssymb}
\usepackage{pifont}
\newcommand{\cmark}{\ding{51}}
\newcommand{\xmark}{\ding{55}}

\usepackage{algorithmic}
\usepackage{array}
\usepackage{booktabs}
\usepackage{multirow}
\usepackage{tabularx}
\usepackage{verbatim}

\usepackage{graphicx}
\usepackage{textcomp}
\usepackage{stfloats}
\usepackage{url}
\usepackage{balance}
\usepackage[most]{tcolorbox}

\usepackage[caption=false,font=normalsize,labelfont=sf,textfont=sf]{subfig}

\def\BibTeX{{\rm B\kern-.05em{\sc i\kern-.025em b}\kern-.08em
		T\kern-.1667em\lower.7ex\hbox{E}\kern-.125emX}}

\newtcolorbox{eadrobox}{
	colback=gray!8,
	colframe=gray!35,
	boxrule=0.3pt,
	arc=1mm,
	left=1.5mm,
	right=1.5mm,
	top=0.8mm,
	bottom=0.8mm,
	fontupper=\centering,
	before skip=0pt,
	after skip=0pt
}

\newtcolorbox{oredgebox}{
	colback=blue!5,
	colframe=blue!40,
	boxrule=0.3pt,
	arc=1mm,
	left=1.5mm,
	right=1.5mm,
	top=0.8mm,
	bottom=0.8mm,
	fontupper=\centering,
	before skip=0pt,
	after skip=0pt
}

\usepackage{hyperref}

\begin{document}
	
	\title{OrEdge: Efficient Multi-Modal Anomaly Detection in Distributed Software Systems via Orthogonal-Domain Learning}
	
	\author{
		Amr M. Zaki,
		Farhoud Jafari Kaleibar,
		Honggeun Ji,
		Komal Sarda,
		Marin Litoiu%
		\thanks{
			Amr M. Zaki (corresponding author), Farhoud Jafari Kaleibar, Honggeun Ji, Komal Sarda, and Marin Litoiu are with the Lassonde School of Engineering, York University, Toronto, ON, Canada.
			(e-mail: amrzaki@yorku.ca; farhoud@yorku.ca; gihong96@yorku.ca; komal253@yorku.ca; mlitoiu@yorku.ca).
		}
	}
	
	\maketitle
	
	\begin{abstract}
		We introduce Orthogonal-Edge (OrEdge), a lightweight framework for real-time anomaly detection in multi-modal distributed software systems. Unlike existing approaches that rely on computationally expensive attention- and graph-based architectures, OrEdge leverages orthogonal-domain temporal representations to achieve accurate anomaly detection with substantially lower computational complexity and model size. It jointly analyzes heterogeneous monitoring data, including logs, metrics, and traces, to identify abnormal software behavior, capture temporal dependencies, and reduce redundancy across observability signals. At its core, OrEdge incorporates OrEdgeCore, a lightweight orthogonal-domain reconstruction module that captures recurring temporal patterns while suppressing transient variations.
		Evaluated on three real-world microservice datasets (MSDS, SN, and TT), OrEdge achieves competitive detection performance while reducing the reconstruction model size to at most 9.6K parameters, compared with 20K--143K parameters in existing methods. This compact design enables efficient deployment on resource-constrained edge devices: on Raspberry Pi platforms, OrEdge achieves sub-second inference and reduces inference latency by over an order of magnitude compared with existing approaches. Extensive ablation studies, sensitivity analyses, orthogonal basis evaluations, and qualitative case studies further validate the effectiveness of each design component.
		Overall, OrEdge demonstrates that orthogonal-domain temporal modeling provides an effective alternative to computationally intensive attention- and graph-based architectures, achieving a favorable balance between detection accuracy and computational efficiency for real-time multi-modal anomaly detection in edge environments. The code is available at \url{https://github.com/theamrzaki/MicroService_Twin_Original}.
	\end{abstract}
	
	\section{Introduction}

The widespread adoption of distributed computing and microservices has enabled scalable, flexible, and resilient software systems by decomposing applications into loosely coupled, independently deployable services~\cite{269-AnoFusion,270-MSTGAD,271-withLLM}. However, this architectural shift also increases operational complexity, making timely anomaly detection essential for maintaining system reliability and availability~\cite{293-microservices-challenges}. Anomalies caused by software defects, hardware failures, network disruptions, or security incidents can rapidly propagate across service dependencies if left undetected~\cite{283-BARO}. As edge computing continues to expand across IoT devices~\cite{HebaEdge}, connected vehicles~\cite{AmrEdge,FarhoudEdge,113-VECInfocom}, UAVs~\cite{151-UAV-Infocom}, and mobile systems~\cite{SherifEdge}, anomaly detection algorithms must provide high detection accuracy while operating under strict computational and memory constraints.

Recent multi-modal anomaly detection approaches~\cite{269-AnoFusion,Eadro,270-MSTGAD,271-withLLM,Art,MUAD,263-complexJournal} improve detection accuracy by jointly modeling logs, metrics, and traces using graph neural networks, Transformers, and large language models. However, these architectures introduce substantial computational and memory overhead, limiting their deployment in real-time and resource-constrained environments. 
Orthogonal-domain methods~\cite{276-Fits,277-filternet,278-FourierGNN,Olinear,280-FreqTimeLoss} provide a promising direction for efficient temporal modeling by learning compact representations with reduced redundancy. Nevertheless, existing approaches mainly target generic or single-source time-series and remain largely unexplored for multi-modal software observability. 
Conversely, lightweight anomaly detectors~\cite{FastRaspberryAnomaly-Stat-1-348,FastRaspberryAnomaly-Neural-2-347} enable efficient edge deployment but are limited to individual telemetry sources, preventing them from capturing cross-modal dependencies among heterogeneous monitoring signals. Therefore, developing an accurate multi-modal anomaly detection framework that maintains low computational cost for edge deployment remains an open challenge.

To address these limitations, we propose \textbf{Orthogonal-Edge (OrEdge)}, a lightweight framework for multi-modal anomaly detection in distributed software systems. OrEdge jointly analyzes logs, metrics, and traces by projecting heterogeneous monitoring signals into orthogonal temporal representations that reduce redundancy while preserving informative temporal patterns. At its core, OrEdge introduces \textbf{OrEdgeCore}, a lightweight orthogonal reconstruction module that models recurring temporal behavior using linear projections and linear attention instead of computationally expensive Transformer or graph-attention architectures. This design enables efficient multi-modal anomaly detection while maintaining low computational and memory overhead.

We evaluate OrEdge on three real-world microservice datasets (SN, TT, and MSDS) against representative anomaly detection baselines. OrEdge achieves competitive detection performance while requiring only up to 9.6K reconstruction parameters, substantially reducing model complexity compared with existing graph-based and multi-modal anomaly detection approaches. Furthermore, OrEdge enables practical deployment on resource-constrained edge platforms, including Raspberry Pi 3 and Raspberry Pi 5, achieving sub-second inference and reducing latency by over an order of magnitude compared with existing methods. These results demonstrate that orthogonal-domain temporal modeling provides an effective alternative to computationally intensive graph- and attention-based architectures for real-time multi-modal anomaly detection.

This work is guided by the following research questions:
\begin{itemize}
	\item \textbf{RQ1:} How does OrEdge compare with state-of-the-art multi-modal anomaly detection approaches in terms of accuracy?
	
	\item \textbf{RQ2:} How does OrEdge enable efficient edge deployment in terms of model complexity and inference latency?
	
	\item \textbf{RQ3:} What are the contributions of OrEdge components and orthogonal projection spaces to anomaly detection performance?
	
	\item \textbf{RQ4:} How sensitive is OrEdge to key design parameters across different multi-modal workloads?
	
	\item \textbf{RQ5:} A qualitative analysis of OrEdge's anomaly detection behavior compared with existing methods.
\end{itemize}

\textbf{The main contributions of this work are summarized as follows:}
\begin{itemize}
	\item We propose \textbf{OrEdge}, a lightweight orthogonal-domain framework for multi-modal anomaly detection that jointly models logs, metrics, and traces through compact temporal representations.
	
	\item We introduce \textbf{OrEdgeCore}, an orthogonal reconstruction module that substantially reduces model complexity while preserving competitive anomaly detection performance.
	
	\item We conduct comprehensive experiments on three real-world microservice datasets, evaluating detection accuracy, computational efficiency, orthogonal basis selection, architectural ablations, parameter sensitivity, and deployment performance.
	
	\item We demonstrate practical edge deployment on Raspberry Pi platforms, showing substantial reductions in latency.
\end{itemize}

\begin{table*}[t]
	\small
	\centering
	\renewcommand{\arraystretch}{1.15}
	\setlength{\tabcolsep}{5pt}
	
	\caption{Comparison of Anomaly Detection Approaches for Multi-Modal Data}
	\label{tab:related_works_comparison}
	
	\begin{tabular}{|p{3.2cm}|p{1.8cm}|p{3.0cm}|p{3.0cm}|p{3.0cm}|p{1.5cm}|}
		\hline
		Method/Year & Telemetry & Dependency Modeling & Architecture & Edge Evaluation & Domain \\
		\hline
		
		\multicolumn{6}{|l|}{\cellcolor{gray} Classical and Lightweight Methods} \\\cline{2-6}
		& \multicolumn{5}{|l|}{\cellcolor{lightgray} Single-modality statistical or neural approaches} \\ \hline
		
		\cite{6of271-single-modality} (DeepLog 2017) & Logs & \xmark & LSTM & -- & Time \\ \hline
		\cite{266-AlertRank} (AlertRank 2022) & L+M+A & \xmark & XGBoost & -- & Time \\ \hline
		\cite{FastRaspberryAnomaly-Stat-1-348} (FADSD 2025) & Single TS & \xmark & Spectrum Score & RPi 4B (2GB), STM32 ($<$1MB SRAM) & Freq. \\ \hline
		\cite{FastRaspberryAnomaly-Neural-2-347} (LFTSAD 2025) & Single TS & \xmark & Dual-branch MLP & RPi 4B (2GB), Jetson NX (8GB) & Time \\ \hline

		\multicolumn{6}{|l|}{\cellcolor{gray} Multi-Modal Methods} \\\cline{2-6}
		& \multicolumn{5}{|l|}{\cellcolor{lightgray} Graph-based Fusion Methods} \\ \hline
		
		\cite{269-AnoFusion} (AnoFusion 2023) & L+M+T & \cmark (GAT Graph) & GAT & -- & Time \\ \hline
		\cite{Eadro} (Eadro 2023) & L+M+T & \cmark (Service Graph) & GAT & -- & Time \\ \hline
		\cite{MUAD} (MUAD 2026) & L+M+T & \cmark (Dynamic Graph) & GAT + Confidence Net (Supervised) & -- & Time \\ \hline
		\cite{service-computing-GrassmannManifolds-code-missing} (MGFusion 2026) & L+M+T & \cmark (Grassmann Graph) & Graph Attention & -- & Grassmann \\ \hline

		& \multicolumn{5}{|l|}{\cellcolor{lightgray} Transformer and Cross-Modal Fusion Methods} \\ \hline
		
		\cite{270-MSTGAD} (MSTGAD 2023) & L+M+T & \cmark (Trace Graph) & Spatial-Temporal Transformer & -- & Time \\ \hline
		\cite{271-withLLM} (LLM4MST 2024) & L+M+T & \cmark (Trace Graph) & GPT-2 & -- & Time \\ \hline
		\cite{Art} (Art 2024) & L+M+T & \xmark & Transformer + GRU & -- & Time \\ \hline
		\cite{medicine} (Medicine 2024) & L+M+T & \xmark & BERT + Attention & -- & Time \\ \hline
		\cite{263-complexJournal} (MAD-CMC 2025) & L+M & \cmark (FC Graph) & Cross-modal LSTM & -- & Time \\ \hline

		\multicolumn{6}{|l|}{\cellcolor{gray} Frequency / Orthogonal Domain Methods} \\\cline{2-6}
		& \multicolumn{5}{|l|}{\cellcolor{lightgray} Compact forecasting and reconstruction models} \\ \hline
		
		\cite{278-FourierGNN} (FourierGNN 2023) & Single TS & \cmark (FC Graph) & Linear Layers & -- & Freq. \\ \hline
		\cite{276-Fits,277-filternet} (FITS/FilterNet 2024) & Single TS & \xmark & Linear Layers & -- & Freq. \\ \hline
		\cite{280-FreqTimeLoss} (FreDF 2025) & Single TS & \xmark & Linear Layers & -- & Freq.+Time \\ \hline
		\cite{Olinear} (OLinear 2025) & Single TS & \xmark & NormLinear & -- & Orthogonal \\ \hline
		\cite{279-metriclogsFourierGNN} (FFAD 2025) & L+M & \cmark (Fourier Graph) & Linear Layers & -- & Freq. \\ \hline
		
		\rowcolor{green!20}
		\textbf{OrEdge (Ours)} & \textbf{L+M+T} & \textbf{\xmark (Flattened Graph)} & \textbf{Linear Projection + Linear Attention + NormLinear} & \textbf{RPi 3 (1GB), RPi 5 (16GB)} & \textbf{Orthogonal} \\ \hline
		
	\end{tabular}
	
	\begin{flushleft}
		\tiny
		L: Logs, M: Metrics, T: Traces, A: Alerts, TS: Time Series, FC: Fully Connected.
	\end{flushleft}
	
\end{table*}
	\section{Related Works}

\subsection{Anomaly Detection for Multi-Modal Data}

Anomaly detection is essential for maintaining reliability in complex microservice and distributed systems~\cite{270-MSTGAD,271-withLLM}. Early approaches relied on single-source telemetry~\cite{6of271-single-modality,singlemodality-7-of-271}, but lacked the contextual information required to capture cross-service dependencies and cascading failures~\cite{271-withLLM}. Lightweight approaches such as FADSD~\cite{FastRaspberryAnomaly-Stat-1-348} and LFTSAD~\cite{FastRaspberryAnomaly-Neural-2-347} enable efficient deployment on constrained devices, including Raspberry Pi, NVIDIA Jetson Xavier, and STM32 platforms. However, they operate on single-modality time-series data and cannot exploit correlations among heterogeneous observability signals.
Recent multi-modal approaches~\cite{266-AlertRank,269-AnoFusion,Eadro,MUAD,service-computing-GrassmannManifolds-code-missing,270-MSTGAD,271-withLLM,Art,medicine} jointly analyze logs, metrics, and traces to improve anomaly detection and fault localization. Graph-based methods model service dependencies using graph attention or geometric representations, while Transformer-based approaches employ attention mechanisms and pretrained models to capture complex temporal interactions. However, these architectures introduce substantial computational and memory overhead due to graph construction, attention operations, and large sequence models, limiting their applicability for real-time and resource-constrained deployment.
Consequently, existing approaches exhibit a fundamental trade-off: lightweight models provide efficient deployment but fail to capture cross-modal dependencies, whereas multi-modal frameworks achieve richer anomaly characterization at the cost of increased complexity and reduced deployment efficiency.

\subsection{Orthogonal Domain for Anomaly Detection}

Orthogonal transformations have recently gained attention as an efficient strategy for compact time-series representation learning~\cite{Dlinear,276-Fits,277-filternet,278-FourierGNN,280-FreqTimeLoss,Olinear}. By projecting signals into orthogonal spaces, these approaches reduce redundancy and capture dominant temporal patterns using substantially fewer parameters than attention-based architectures.
FITS~\cite{276-Fits} demonstrated that frequency-domain projections can achieve competitive forecasting performance with only 10,000 parameters compared with hundreds of millions in Transformer-based models. Subsequent studies extended this idea through improved frequency filtering (FilterNet~\cite{277-filternet}), spectral graph modeling (FourierGNN~\cite{278-FourierGNN}), and frequency-time objectives (FreqTimeLoss~\cite{280-FreqTimeLoss}). FourierGNN was later applied to microservice anomaly detection~\cite{279-metriclogsFourierGNN}; however, it considers only logs and metrics and relies on fully connected graphs, while deployment efficiency remains unexplored.
More recently, OLinear~\cite{Olinear} introduced dataset-specific orthogonal bases through temporal correlation analysis. Despite their efficiency advantages, existing orthogonal-domain approaches primarily target generic forecasting or limited telemetry sources and have not been investigated for multi-modal software observability or edge deployment scenarios.

\subsection{Gap and Motivation}

Existing studies reveal two key limitations. First, lightweight anomaly detectors enable efficient edge deployment but are restricted to single-modality telemetry, while multi-modal approaches improve anomaly characterization through graph and attention-based modeling at the cost of significant computational and memory overhead. Second, although orthogonal-domain methods provide compact temporal representations, they have rarely been explored for multi-modal software observability involving logs, metrics, and traces.
To address these challenges, we propose OrEdge, a lightweight semi-supervised framework that integrates multi-modal telemetry with orthogonal-domain learning for efficient anomaly detection. By projecting heterogeneous monitoring signals into compact orthogonal representations, OrEdge reduces redundancy, captures temporal dependencies, and avoids expensive graph construction and large Transformer architectures. This design enables accurate and practical anomaly detection on resource-constrained edge platforms.

	\section{Problem Formulation}

We consider a multi-modal software monitoring scenario deployed on resource-constrained edge devices, where heterogeneous telemetry data, including metrics, logs, and traces, are collected from a set of services over time. Following the semi-supervised formulation adopted in MSTGAD~\cite{270-MSTGAD}, only a subset of service states may be labeled, while the remaining observations are treated as unlabeled. Given a monitoring window of length $T$, the input at time step $t$ is represented as:
\begin{equation}
	X_t = \{M_{t-T+1:t}, L_{t-T+1:t}, S_{t-T+1:t}\},
	\label{eq:input}
\end{equation}
where $M \in \mathbb{R}^{N \times D_m}$, $L \in \mathbb{R}^{N \times D_l}$, and $S \in \mathbb{R}^{N \times N \times D_s}$ denote metrics, logs, and traces collected from $N$ services. Each service is associated with a label $y_t \in \{-1,0,1\}^{N}$, where $1$, $0$, and $-1$ represent anomalous, normal, and unlabeled states, respectively.

The objective is to accurately identify anomalous services while minimizing computational overhead. Following a semi-supervised anomaly detection formulation, the model jointly learns to reconstruct monitoring observations and classify service states by optimizing:
\begin{equation}
	\min_{\theta}
	\mathcal{L}_{\text{recon}}(X_t,\hat{X}_t)
	+ \lambda\mathcal{L}_{\text{cls}}(y_t,\hat{y}_t),
	\label{eq:joint_loss}
\end{equation}
where $\hat{X}_t=F_{\theta}(X_t)$ denotes the reconstructed telemetry data, $\hat{y}_t=G_{\theta}(X_t)$ represents the predicted anomaly states,
$\theta$ are the model parameters, $\mathcal{L}_{\text{recon}}$ is the reconstruction loss, $\mathcal{L}_{\text{cls}}$ is the classification loss,
and $\lambda$ controls the trade-off between reconstruction and classification objectives.

The main challenges arise from the heterogeneous nature of multi-modal telemetry, complex temporal dependencies, and strict resource constraints of edge deployment. Therefore, an effective solution must jointly exploit information from different monitoring sources while maintaining low computational and memory overhead.
	\section{OrEdge Framework}

\begin{figure*}[ht!] 
	\centering
	\includegraphics[width=\textwidth]{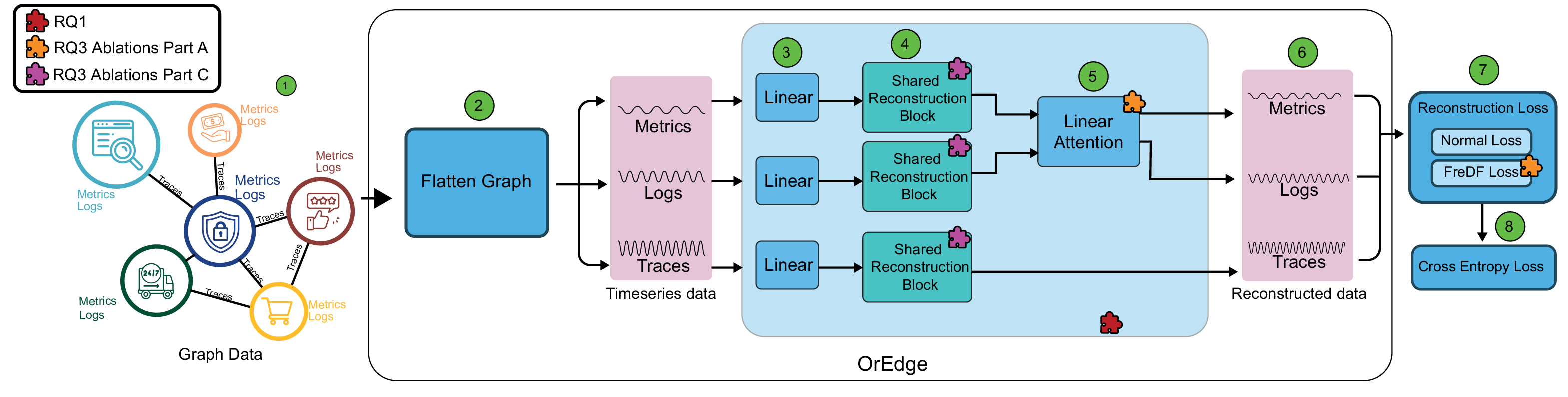}
	\caption{OrEdge framework overview. OrEdge receives a service dependency graph with multi-modal observability data (metrics, logs, traces). The shared reconstruction block captures temporal dependencies and cross-modal interactions, followed by classification to identify anomalies.}
	\label{fig:OrEdge}
\end{figure*}

\subsection{Overview}

OrEdge detects software anomalies by jointly modeling metrics $M$, logs $L$, and traces $T$ through orthogonal-domain representations. Raw observability data often contain redundant temporal patterns and strong correlations across modalities, which increase the difficulty of identifying abnormal behaviors. By projecting heterogeneous telemetry into orthogonal latent spaces,

\[
(M,L,T) \rightarrow (M',L',T'),
\]

OrEdge reduces redundant correlations and extracts compact representations that preserve informative temporal and structural patterns.

As shown in Fig.~\ref{fig:OrEdge}, OrEdge first receives a service dependency graph from trace information. The multi-modal telemetry is then transformed into the orthogonal domain, where lightweight reconstruction and classification modules identify deviations from normal behavior. By exploiting orthogonal representations instead of directly modeling raw time-domain signals, OrEdge enables accurate and efficient anomaly detection for large-scale microservice systems under edge deployment constraints.

\subsection{Pre-Processing}

\paragraph{Metrics and Logs}
Metrics and logs provide complementary views of service behavior. Metrics are represented as $M_t \in \mathbb{R}^{N \times D_m}$, while logs are represented as $L_t \in \mathbb{R}^{N \times D_l}$, where $N$ denotes the number of services. Logs are first parsed into templates and aggregated at the service level.

Both modalities are normalized using min-max scaling:
\begin{equation}
	X_t = \frac{X_t-\min(T)}{\max(T)-\min(T)+\epsilon},
	\quad X_t \in \{M_t,L_t\}.
\end{equation}

\paragraph{Traces}
Traces capture service dependencies through span-level information, including interaction frequency, latency, and duration statistics. Aggregating trace information within each monitoring window produces:
\begin{equation}
	S_t \in \mathbb{R}^{N \times N \times D_s}.
\end{equation}

Each entry in $S_t$ represents the interaction features between a pair of services during the corresponding time window. To reduce scale variations across different workloads, trace features are normalized as:
\begin{equation}
	S_t=\frac{S_t}{\operatorname{mean}(T_{\text{Trace}})}.
\end{equation}
\paragraph{Graph Data}

OrEdge first receives a multimodal service dependency graph:
\begin{equation}
	\mathcal{G}_t=\langle V_t,A_t,E_t\rangle,
\end{equation}
where nodes represent microservices and edges represent service interactions extracted from distributed traces (Fig.~\ref{fig:OrEdge}).

Node features combine service-level metrics and logs:
\begin{equation}
	V_t=M_t \parallel L_t,
\end{equation}
while edge attributes are obtained from trace information:
\begin{equation}
	E_t=S_t,
\end{equation}
where $S_t$ contains interaction statistics such as invocation frequency and latency. This representation unifies heterogeneous telemetry while preserving service dependencies.

\paragraph{Graph Flattening and Temporal Modeling}

To capture temporal patterns, OrEdge groups consecutive service graphs into sliding windows 
$\mathcal{D}_t=\{\mathcal{G}_{t-k+1},\ldots,\mathcal{G}_t\}$, where $k$ is the window length. 
Each window is associated with service-level labels $y_t\in\mathbb{R}^{1\times N}$ indicating whether each service is abnormal ($1$), normal ($0$), or unknown ($-1$).

Rather than using graph neural networks to perform message passing, OrEdge converts the graph data into temporal sequences that can be processed by lightweight temporal models. The metric and log features of each service are arranged into sequences $\widetilde{\mathbf{X}}^{(m)}\in\mathbb{R}^{BN\times T\times F_m}$ and $\widetilde{\mathbf{X}}^{(l)}\in\mathbb{R}^{BN\times T\times F_l}$, while trace features between interacting services are organized as $\widetilde{\mathbf{X}}^{(s)}\in\mathbb{R}^{BE\times T\times F_s}$. Here, $B$ is the batch size, $N$ is the number of services, $E$ is the number of service interactions, $T$ is the temporal window length, and $F_m$, $F_l$, and $F_s$ are the corresponding feature dimensions.

This transformation preserves both service behavior and service interaction patterns over time while avoiding the computational overhead of graph message passing.
\subsection{Multi-Modal Reconstruction Block}

OrEdge reconstructs heterogeneous monitoring signals, including metrics, logs, and traces, through a unified multi-modal reconstruction framework. Unlike conventional approaches that independently process each modality or rely on computationally expensive graph-based fusion, OrEdge separates modality-specific representation learning from shared temporal reconstruction while employing modality-aware fusion strategies.

Existing lightweight temporal models, including frequency-domain forecasting and linear sequence models, are primarily designed for single-stream prediction tasks. Directly applying these approaches to multi-modal observability data is challenging because different telemetry sources exhibit distinct temporal characteristics and anomaly manifestations. Moreover, distributed system anomalies often appear as inconsistencies between multiple signals rather than isolated deviations within a single modality. Therefore, OrEdge integrates lightweight temporal modeling into a reconstruction framework that preserves modality-specific characteristics while enabling efficient interaction between complementary observability sources.

Given input modalities $\mathcal{X}^{(m)}$, $\mathcal{X}^{(l)}$, and $\mathcal{X}^{(t)}$, each modality is first projected into a latent representation through lightweight modality-specific encoders:

\begin{equation}
	Z^{(m)}=\phi_m(\mathcal{X}^{(m)}), \quad
	Z^{(l)}=\phi_l(\mathcal{X}^{(l)}), \quad
	Z^{(t)}=\phi_t(\mathcal{X}^{(t)}),
\end{equation}

where $\phi_m$, $\phi_l$, and $\phi_t$ are linear projection layers that adapt heterogeneous input dimensions while preserving modality-specific characteristics.

The projected representations are then processed by a shared temporal reconstruction module:

\begin{equation}
	H^{(m)}=f(Z^{(m)}), \quad
	H^{(l)}=f(Z^{(l)}), \quad
	H^{(t)}=f(Z^{(t)}),
\end{equation}

where $f(\cdot)$ captures temporal dependencies within each modality. Within this framework, OrEdge instantiates $f(\cdot)$ using OrEdgeCore, which adapts orthogonal-domain temporal modeling, constrained linear transformation, and temporal filtering into an efficient reconstruction mechanism for multi-modal anomaly detection. Rather than introducing a new temporal operator in isolation, OrEdgeCore integrates these lightweight components into a shared reconstruction architecture tailored for heterogeneous system telemetry.

To model interactions between dense temporal observability signals, OrEdge applies lightweight linear attention between metrics and logs:

\begin{equation}
	\widetilde{H}^{(m,l)}
	=
	\mathrm{LinearAttn}(H^{(m)},H^{(l)}),
\end{equation}

which captures metric-log dependencies with substantially lower computational cost than full attention mechanisms. In contrast, trace information is maintained as structural dependency features:

\begin{equation}
	H^{(t)}=f(Z^{(t)}),
\end{equation}

allowing OrEdge to preserve service relationship information without requiring computationally expensive graph message passing.

The fused metric-log representation is subsequently separated into modality-specific representations:

\begin{equation}
	\widetilde{H}^{(m)},\widetilde{H}^{(l)}
	=
	\mathrm{Split}(\widetilde{H}^{(m,l)}),
\end{equation}

and combined with the independently reconstructed trace representation through modality-specific reconstruction heads.

This modality-aware design enables OrEdge to exploit complementary information across heterogeneous observability sources: metric-log interactions provide temporal correlation modeling, while trace reconstruction preserves service dependency information. Consequently, OrEdge transforms lightweight temporal modeling techniques originally developed for forecasting into an efficient reconstruction-based anomaly detection framework for distributed software systems.

\subsection{\textbf{OrEdgeCore Shared Model}}  
\label{sec:OrEdgeCore}

\begin{figure}[ht!] 
	\centering
	\includegraphics[width=\linewidth]{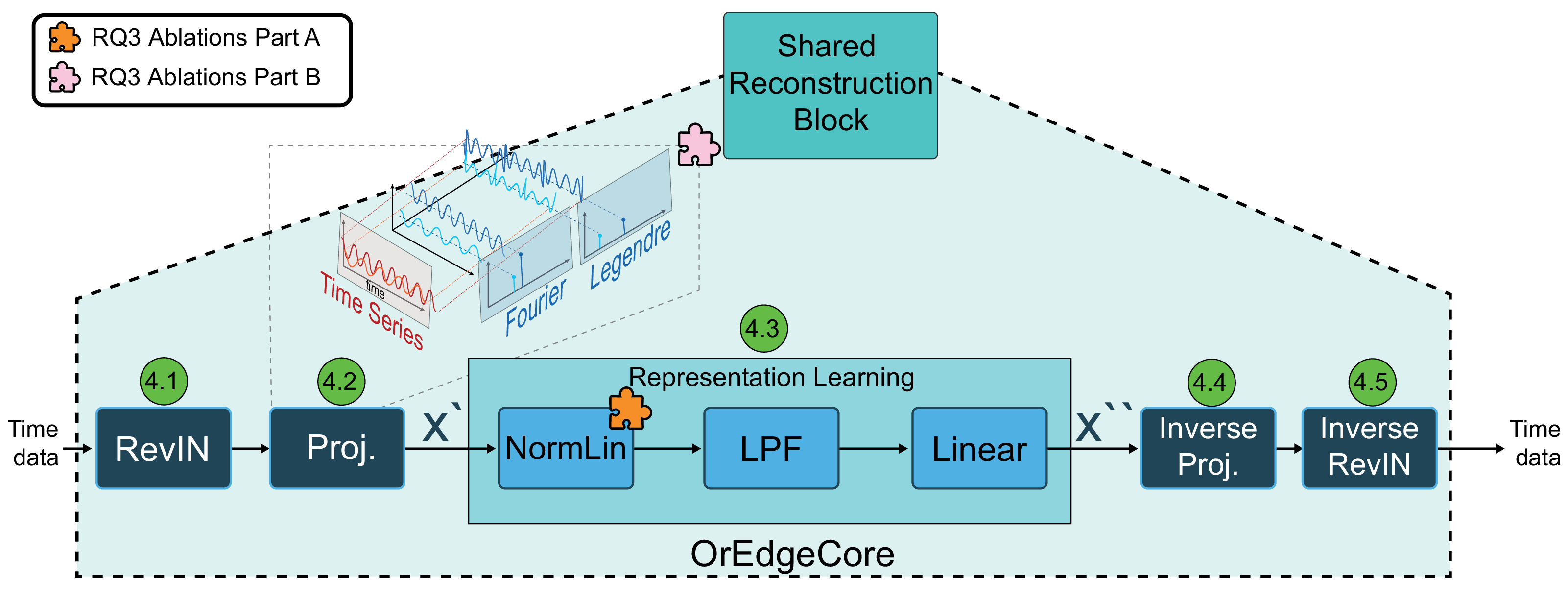}
	\caption{OrEdgeCore shared reconstruction block. OrEdgeCore projects multi-modal monitoring data into an orthogonal domain, applies a constrained linear layer to learn stable representations, and uses a low-pass filter to reduce noise before mapping back to the time domain.}
	\label{fig:OrEdgeCore}
\end{figure}
In addition to existing reconstruction backbones, we propose OrEdgeCore, an orthogonal-domain temporal model designed to efficiently reconstruct multi-modal monitoring data while capturing long-term dependencies and reducing temporal redundancy.

Given an input series $X$ from metrics, logs, or traces, OrEdgeCore first applies reversible normalization using RevIN~\cite{kim2022reversible} to reduce non-stationarity. The normalized representation is then projected into an orthogonal domain:
\begin{equation}
	\begin{aligned}
		x' &= \mathrm{Projection}(\mathrm{RevIN}(X)),\\
		\mathrm{Projection} &\in \{\text{Cheb, Fourier, Legendre, Laguerre, Hermite}\}.
	\end{aligned}
\end{equation}
This transformation separates temporal patterns into compact components, enabling efficient representation learning.

Next, OrEdgeCore applies NormLin~\cite{Olinear} to learn stable feature interactions, followed by low-pass filtering and linear projection:
\begin{equation}
	x''=\mathrm{Linear}(\mathrm{LPF}(\mathrm{NormLin}(x'))).
\end{equation}
The reconstructed signal is then mapped back to the original temporal domain using the inverse orthogonal projection and RevIN transformation.

By combining orthogonal representations, constrained linear modeling, and temporal filtering, OrEdgeCore provides a lightweight reconstruction mechanism that preserves informative temporal dependencies while reducing redundancy. This enables efficient anomaly detection across heterogeneous observability data without relying on computationally expensive sequence models.
\subsection{Loss Calculation}

OrEdge follows the semi-supervised formulation of MSTGAD~\cite{270-MSTGAD}, jointly optimizing reconstruction and classification objectives. Reconstruction provides the primary learning signal, while classification exploits available anomaly labels.

\paragraph{Orthogonal Reconstruction Loss}
To preserve temporal patterns beyond the time domain, OrEdge incorporates an orthogonal-domain reconstruction objective using FreDF~\cite{280-FreqTimeLoss}:

\begin{equation}
	\mathcal{L}_{\text{recon}} =
	\frac{1}{n} \sum_{i=1}^{n} 
	\underbrace{\| X_i - \hat{X}_i \|_2^2}_{\text{time-domain}}
	+
	\lambda 
	\underbrace{\| \mathcal{O}(\hat{X}_i)-\mathcal{O}(X_i) \|_1}_{\text{orthogonal-domain (FreDF)}},
\end{equation}

where $\mathcal{O}(\cdot)$ denotes the orthogonal transformation and $\lambda$ controls the contribution of the FreDF objective.

\paragraph{Semi-Supervised Classification Loss}
Following MSTGAD~\cite{270-MSTGAD}, a weighted cross-entropy loss is added when anomaly labels are available:

\begin{equation}
	\mathcal{L}_{\text{cls}} =
	-\frac{1}{n}\sum_{i=1}^{n}\sum_{c\in C}
	w_c y_{i,c}\log(\hat{y}_{i,c}),
\end{equation}

where $w_c$ addresses class imbalance. The final objective is:

\begin{equation}
	\mathcal{L}=
	\mathcal{L}_{\text{recon}}
	+\alpha\mathcal{L}_{\text{cls}},
\end{equation}

where $\alpha$ balances reconstruction and classification objectives.

	\section{EVALUATION}

\subsection{Datasets}

We evaluate OrEdge on three multi-modal microservice datasets: MSDS, Social Network (SN), and Train Ticket (TT), which contain logs, metrics, and traces as summarized in Table~\ref{tab:datasets}.

\textbf{MSDS}~\cite{msds-42-of-270} is collected from an OpenStack-based distributed AI analytics system and includes trace information, application logs, and system metrics from five nodes. It provides fine-grained anomaly annotations with anomaly types and temporal boundaries.

\textbf{SN}~\cite{SN-42-of-Eadro} represents a social networking system consisting of 21 microservices (14 business-related) communicating through Thrift RPC. It contains multi-modal observability data, including logs, metrics, and traces.

\textbf{TT}~\cite{TT-17-of-Eadro} is an open-source railway ticketing benchmark with 41 interacting microservices (27 business-related). Similar to SN, it provides logs, metrics, and traces for monitoring service behavior.

\begin{table}[t]
\centering
\caption{Dataset Summary}
\label{tab:datasets}
\setlength{\tabcolsep}{3pt}
\renewcommand{\arraystretch}{0.95}
\begin{tabular}{lccc}
\toprule
\scriptsize
\textbf{Data} & \textbf{Services} & \textbf{Modalities} & \textbf{Anomaly Types} \\
\midrule
MSDS & 5  & L/M/T &  Injected (labeled) \\
SN   & 21 & L/M/T & Behavior, Performance, Intrusion \\
TT   & 41 & L/M/T & Behavior, Performance, Network \\
\bottomrule
\end{tabular}
\end{table}

\subsection{Experimental Setup}

\paragraph{\textbf{Baselines and Comparison:}}
We evaluate OrEdge against representative anomaly detection approaches to assess its detection capability, efficiency, and generalizability.

For multi-modal anomaly detection, we compare OrEdge with recent state-of-the-art methods, including Eadro~\cite{Eadro}, AnoFusion~\cite{269-AnoFusion}, MSTGAD~\cite{270-MSTGAD}, Art~\cite{Art}, and Medicine~\cite{medicine}. These approaches represent diverse modeling paradigms, including graph-based fusion, attention mechanisms, and advanced temporal representation learning for jointly analyzing heterogeneous observability data. All baselines are re-implemented using publicly available implementations to ensure a fair and consistent comparison.
Recent multi-modal anomaly detection approaches, including MUAD~\cite{MUAD} and MGFusion~\cite{service-computing-GrassmannManifolds-code-missing}, are not included in the main quantitative comparison due to methodological and reproducibility differences. MUAD formulates anomaly detection as a fully supervised classification problem, requiring anomaly labels during training, whereas OrEdge follows a semi-supervised reconstruction-based paradigm designed for scenarios with limited anomaly annotations. Therefore, a direct comparison under the original settings would not provide an equivalent evaluation. MGFusion is excluded because its implementation is not publicly available, preventing a reproducible evaluation under our experimental framework.
To evaluate the generalizability of the proposed framework, we further replace the OrEdge temporal core with alternative reconstruction backbones, including iTransformer~\cite{itransformer} and FEDformer~\cite{fedformer}. This analysis examines whether the proposed orthogonal-domain representation remains effective across different temporal modeling architectures rather than being tied to a specific backbone.
All experiments are repeated with multiple random seeds to ensure statistical robustness, and the mean and standard deviation of the results are reported.

\paragraph{\textbf{Evaluation Metrics:}}
We evaluate OrEdge from two complementary perspectives: detection performance and computational efficiency.

\textbf{Accuracy metrics.}
We formulate anomaly detection as a point-wise binary classification problem~\cite{270-MSTGAD} and report standard metrics, including precision, recall, and F1-score, following previous studies~\cite{Eadro,269-AnoFusion,Art}. During inference, OrEdge outputs a two-class probability distribution through a softmax classification layer, and the predicted labels are obtained using the argmax operation. Therefore, all classification metrics are computed directly from point-wise predictions without point adjustment, event-level post-processing, or test-time threshold optimization. Additionally, we report threshold-independent metrics, including Area Under the ROC Curve (AUC-ROC) and Average Precision (AP)~\cite{msds-42-of-270}, to provide a comprehensive evaluation under class-imbalanced anomaly distributions.

\textbf{Efficiency metrics.}
We evaluate computational efficiency in terms of model complexity and runtime performance. Specifically, we report the number of trainable parameters in the reconstruction module, following previous lightweight time-series modeling studies~\cite{276-Fits}, together with FLOPs required for a single forward pass and server-side training time. For edge deployment evaluation, we follow the methodology of~\cite{FastRaspberryAnomaly-Neural-2-347,FastRaspberryAnomaly-Stat-1-348} in which we measure per-sample inference latency and peak memory usage on resource-constrained devices.

\paragraph{\textbf{Implementation:}}
For all datasets, we adopt the standard training and testing splits from previous studies, using~\cite{msds-42-of-270} for MSDS and~\cite{Eadro} for SN and TT. To ensure a fair comparison, the embedding dimensions of all input modalities are fixed across models within each dataset. Specifically, MSDS uses embedding dimensions of 4, 4, and 16 for metrics, traces, and logs, respectively, while SN and TT use 8, 8, and 32. The number of attention heads is also kept consistent across transformer-based models.
All models are trained for 300 epochs on MSDS and SN, and 50 epochs on TT due to its larger scale. A temporal window size of 10 is used across all experiments, with a dropout rate of 0.2 to reduce overfitting. Models are optimized using the AdaBelief optimizer with an initial learning rate of 0.001. The linear attention dimension and orthogonal-domain loss weight are selected based on the sensitivity analysis for each dataset. All baseline models are configured using their recommended hyperparameters. The full hyperparameter settings, the trained weights, full csv results, scripts to reproduce the experiments, scripts to generate the figures and tables for all models and links to the open source datasets are provided in the code repository for reproducibility.
All experiments are conducted on a Linux workstation equipped with an Intel(R) Core(TM) i9-10900K CPU @ 3.70GHz (20 cores), 32 GB RAM, and an NVIDIA GeForce RTX 3070 GPU with 8 GB memory, running Ubuntu 22.04.2 LTS. The implementation is developed using Python 3.10.12, PyTorch 2.7.1+cu126 with CUDA 12.6, and PyTorch Geometric 2.6.1.
To evaluate practical edge deployment, we additionally deploy OrEdge on two resource-constrained platforms: Raspberry Pi 5 equipped with an ARM Cortex-A76 processor and 16 GB RAM, and Raspberry Pi 3 equipped with an ARM Cortex-A53 processor and 1 GB RAM. On these devices, we measure inference latency and peak memory usage to assess the feasibility of real-time multi-modal anomaly detection in constrained environments.

\subsection{Results and Analysis}


\subsubsection{\textbf{(RQ1) Detection Performance}}

\noindent As shown in Table~\ref{tab:accuracy_metrics}, different modeling paradigms for this research question exhibit distinct trade-offs in accuracy. Results are reported as averages over multiple random seeds to ensure statistical robustness.
\begin{table}[t]
	\caption{Anomaly Detection Accuracy Performance Across Datasets (Mean $\pm$ SD)}
	\label{tab:accuracy_metrics}
	\centering
	\scriptsize
	\begin{tabular}{p{1.5cm}p{1.3cm}p{1.4cm}p{1.4cm}p{1.4cm}}
		\toprule
		\textbf{Metric} & \textbf{Model Variant} & \textbf{SN} & \textbf{TT} & \textbf{MSDS} \\
		\midrule
		Precision                 & Eadro                & \cellcolor{yellow!0!red!35}0.907 {\scriptsize$\pm$0.016} & \cellcolor{green!95!yellow!35}0.959 {\scriptsize$\pm$0.000} & \cellcolor{yellow!0!red!35}0.877 {\scriptsize$\pm$0.002} \\
		& AnoFusion            & \cellcolor{yellow!27!red!35}0.917 {\scriptsize$\pm$0.010} & \cellcolor{green!99!yellow!35}0.962 {\scriptsize$\pm$0.010} & \cellcolor{green!31!yellow!35}0.931 {\scriptsize$\pm$0.014} \\
		& MSTGAD               & \cellcolor{green!100!yellow!35}0.983 {\scriptsize$\pm$0.004} & \cellcolor{green!36!yellow!35}0.923 {\scriptsize$\pm$0.005} & \cellcolor{green!100!yellow!35}0.959 {\scriptsize$\pm$0.015} \\
		& Art                  & \cellcolor{green!98!yellow!35}0.982 {\scriptsize$\pm$0.004} & \cellcolor{green!100!yellow!35}0.962 {\scriptsize$\pm$0.008} & \cellcolor{yellow!39!red!35}0.893 {\scriptsize$\pm$0.002} \\
		& Medicine             & \cellcolor{green!40!yellow!35}0.960 {\scriptsize$\pm$0.017} & \cellcolor{yellow!0!red!35}0.840 {\scriptsize$\pm$0.095} & \cellcolor{green!75!yellow!35}0.949 {\scriptsize$\pm$0.009} \\
		& OrEdge               & \cellcolor{green!70!yellow!35}0.971 {\scriptsize$\pm$0.005} & \cellcolor{green!70!yellow!35}0.944 {\scriptsize$\pm$0.002} & \cellcolor{green!77!yellow!35}0.950 {\scriptsize$\pm$0.008} \\
		\cmidrule{1-5}
		Recall                    & Eadro                & \cellcolor{yellow!48!red!35}0.899 {\scriptsize$\pm$0.043} & \cellcolor{green!29!yellow!35}0.936 {\scriptsize$\pm$0.002} & \cellcolor{green!24!yellow!35}0.961 {\scriptsize$\pm$0.017} \\
		& AnoFusion            & \cellcolor{yellow!0!red!35}0.872 {\scriptsize$\pm$0.009} & \cellcolor{green!67!yellow!35}0.965 {\scriptsize$\pm$0.006} & \cellcolor{yellow!0!red!35}0.926 {\scriptsize$\pm$0.003} \\
		& MSTGAD               & \cellcolor{green!90!yellow!35}0.979 {\scriptsize$\pm$0.001} & \cellcolor{green!76!yellow!35}0.972 {\scriptsize$\pm$0.008} & \cellcolor{green!44!yellow!35}0.966 {\scriptsize$\pm$0.012} \\
		& Art                  & \cellcolor{green!100!yellow!35}0.984 {\scriptsize$\pm$0.003} & \cellcolor{green!100!yellow!35}0.990 {\scriptsize$\pm$0.002} & \cellcolor{green!62!yellow!35}0.972 {\scriptsize$\pm$0.019} \\
		& Medicine             & \cellcolor{green!88!yellow!35}0.977 {\scriptsize$\pm$0.014} & \cellcolor{yellow!0!red!35}0.838 {\scriptsize$\pm$0.131} & \cellcolor{green!74!yellow!35}0.975 {\scriptsize$\pm$0.030} \\
		& OrEdge               & \cellcolor{green!54!yellow!35}0.958 {\scriptsize$\pm$0.008} & \cellcolor{green!61!yellow!35}0.961 {\scriptsize$\pm$0.007} & \cellcolor{green!100!yellow!35}0.982 {\scriptsize$\pm$0.012} \\
		\cmidrule{1-5}
		Avg Precision             & Eadro                & \cellcolor{yellow!0!red!35}0.949 {\scriptsize$\pm$0.010} & \cellcolor{green!71!yellow!35}0.987 {\scriptsize$\pm$0.001} & \cellcolor{green!100!yellow!35}0.983 {\scriptsize$\pm$0.002} \\
		& AnoFusion            & \cellcolor{green!43!yellow!35}0.984 {\scriptsize$\pm$0.001} & \cellcolor{green!77!yellow!35}0.990 {\scriptsize$\pm$0.005} & \cellcolor{green!42!yellow!35}0.977 {\scriptsize$\pm$0.000} \\
		& MSTGAD               & \cellcolor{green!99!yellow!35}0.997 {\scriptsize$\pm$0.001} & \cellcolor{green!86!yellow!35}0.993 {\scriptsize$\pm$0.000} & \cellcolor{green!67!yellow!35}0.979 {\scriptsize$\pm$0.002} \\
		& Art                  & \cellcolor{green!100!yellow!35}0.997 {\scriptsize$\pm$0.001} & \cellcolor{green!100!yellow!35}0.998 {\scriptsize$\pm$0.000} & \cellcolor{yellow!0!red!35}0.963 {\scriptsize$\pm$0.011} \\
		& Medicine             & \cellcolor{green!77!yellow!35}0.991 {\scriptsize$\pm$0.009} & \cellcolor{yellow!0!red!35}0.919 {\scriptsize$\pm$0.048} & \cellcolor{green!20!yellow!35}0.975 {\scriptsize$\pm$0.011} \\
		& OrEdge               & \cellcolor{green!77!yellow!35}0.992 {\scriptsize$\pm$0.002} & \cellcolor{green!88!yellow!35}0.994 {\scriptsize$\pm$0.001} & \cellcolor{green!54!yellow!35}0.978 {\scriptsize$\pm$0.006} \\
		\cmidrule{1-5}
		F1-Score                  & Eadro                & \cellcolor{yellow!18!red!35}0.902 {\scriptsize$\pm$0.016} & \cellcolor{green!60!yellow!35}0.947 {\scriptsize$\pm$0.001} & \cellcolor{yellow!0!red!35}0.917 {\scriptsize$\pm$0.008} \\
		& AnoFusion            & \cellcolor{yellow!0!red!35}0.894 {\scriptsize$\pm$0.002} & \cellcolor{green!82!yellow!35}0.963 {\scriptsize$\pm$0.002} & \cellcolor{yellow!47!red!35}0.929 {\scriptsize$\pm$0.005} \\
		& MSTGAD               & \cellcolor{green!95!yellow!35}0.981 {\scriptsize$\pm$0.002} & \cellcolor{green!59!yellow!35}0.947 {\scriptsize$\pm$0.003} & \cellcolor{green!87!yellow!35}0.963 {\scriptsize$\pm$0.004} \\
		& Art                  & \cellcolor{green!100!yellow!35}0.983 {\scriptsize$\pm$0.004} & \cellcolor{green!100!yellow!35}0.976 {\scriptsize$\pm$0.004} & \cellcolor{yellow!56!red!35}0.931 {\scriptsize$\pm$0.010} \\
		& Medicine             & \cellcolor{green!67!yellow!35}0.968 {\scriptsize$\pm$0.009} & \cellcolor{yellow!0!red!35}0.834 {\scriptsize$\pm$0.087} & \cellcolor{green!83!yellow!35}0.962 {\scriptsize$\pm$0.011} \\
		& OrEdge               & \cellcolor{green!59!yellow!35}0.965 {\scriptsize$\pm$0.004} & \cellcolor{green!66!yellow!35}0.952 {\scriptsize$\pm$0.003} & \cellcolor{green!100!yellow!35}0.966 {\scriptsize$\pm$0.004} \\
		\cmidrule{1-5}
		AUC-ROC                   & Eadro                & \cellcolor{yellow!0!red!35}0.994 {\scriptsize$\pm$0.001} & \cellcolor{green!69!yellow!35}0.995 {\scriptsize$\pm$0.002} & \cellcolor{green!100!yellow!35}0.999 {\scriptsize$\pm$0.000} \\
		& AnoFusion            & \cellcolor{green!98!yellow!35}0.998 {\scriptsize$\pm$0.000} & \cellcolor{green!74!yellow!35}0.996 {\scriptsize$\pm$0.003} & \cellcolor{yellow!0!red!35}0.986 {\scriptsize$\pm$0.003} \\
		& MSTGAD               & \cellcolor{green!48!yellow!35}0.997 {\scriptsize$\pm$0.001} & \cellcolor{green!95!yellow!35}0.999 {\scriptsize$\pm$0.000} & \cellcolor{yellow!99!red!35}0.993 {\scriptsize$\pm$0.002} \\
		& Art                  & \cellcolor{green!100!yellow!35}0.998 {\scriptsize$\pm$0.001} & \cellcolor{green!100!yellow!35}1.000 {\scriptsize$\pm$0.000} & \cellcolor{green!71!yellow!35}0.997 {\scriptsize$\pm$0.001} \\
		& Medicine             & \cellcolor{green!96!yellow!35}0.998 {\scriptsize$\pm$0.002} & \cellcolor{yellow!0!red!35}0.969 {\scriptsize$\pm$0.034} & \cellcolor{green!53!yellow!35}0.996 {\scriptsize$\pm$0.000} \\
		& OrEdge               & \cellcolor{yellow!1!red!35}0.994 {\scriptsize$\pm$0.001} & \cellcolor{green!97!yellow!35}1.000 {\scriptsize$\pm$0.000} & \cellcolor{green!30!yellow!35}0.994 {\scriptsize$\pm$0.006} \\
		\bottomrule
	\end{tabular}
\end{table}

Table~\ref{tab:accuracy_metrics} compares OrEdge against representative multi-modal anomaly detection methods across the SN, TT, and MSDS datasets using Precision, Recall, Average Precision (AP), F1-score, and AUC-ROC. Overall, OrEdge achieves competitive anomaly detection performance across all datasets while substantially reducing model complexity and computational cost as would be highlighted in RQ2.

\noindent\textbf{Comparison with Multi-Modal Baselines.}
Existing graph-based approaches, such as MSTGAD and Art, achieve the highest F1-scores on SN and TT, benefiting from their explicit graph modeling capabilities. Specifically, Art and MSTGAD obtain F1-scores of 0.983 and 0.981 on SN, while Art achieves 0.976 on TT. In comparison, OrEdge achieves F1-scores of 0.965 and 0.952 on these datasets, representing a small performance trade-off for significantly lower computational overhead.
On MSDS, however, OrEdge achieves the best overall detection performance, obtaining the highest Recall (0.982) and F1-score (0.966), outperforming Eadro (0.917), AnoFusion (0.929), and comparable graph-based methods. These results demonstrate that OrEdge can effectively capture complex multi-modal failure patterns while maintaining a lightweight architecture.

\noindent\textbf{Robustness Across Datasets.}
Despite the differences in absolute performance, OrEdge demonstrates stable behavior across diverse microservice environments. It consistently achieves high Average Precision (0.978--0.994) and AUC-ROC (0.994--1.000), indicating strong anomaly ranking capability independent of the classification threshold.
These results highlight the central advantage of OrEdge: instead of maximizing detection accuracy through expensive graph neural networks or complex fusion mechanisms, it achieves competitive accuracy with substantially lower reconstruction parameters and computational requirements. This favorable accuracy-efficiency trade-off enables practical deployment of multi-modal anomaly detection on resource-constrained edge devices.

\begin{tcolorbox}[
	colback=gray!10,
	colframe=black!50,
	title={RQ1 Summary: Detection Performance Evaluation},
	left=0mm,
	right=1mm,
	top=1mm,
	bottom=1mm,
	boxsep=1mm,
	arc=2mm
	]
	\begin{itemize}
		\item OrEdge achieves competitive performance across all datasets while using a lightweight orthogonal-domain representation without relying on complex architectures. 
		
	\end{itemize}
\end{tcolorbox}


\subsubsection{\textbf{(RQ2) Efficiency Analysis}}

\begin{table*}[t]
	\caption{Efficiency and Resource Consumption Analysis Across Devices (Mean $\pm$ SD)}
	\label{tab:efficiency_metrics}
	\centering
	\scriptsize
	\begin{tabular}{llcccc}
		\toprule
		\textbf{Device} & \textbf{Metric} & \textbf{Model Variant} & \textbf{SN} & \textbf{TT} & \textbf{MSDS} \\
		\midrule
		\smash{\begin{tabular}[t]{l}\textbf{Server}\\ \footnotesize (GPU: RTX 3070 CPU:i9)\end{tabular}} & Training Time (GPU) [s]             & Eadro                & \cellcolor{yellow!2!green!35}1.4 {\scriptsize$\pm$0.0} & \cellcolor{yellow!0!green!35}18.5 {\scriptsize$\pm$0.4} & \cellcolor{yellow!10!green!35}5.2 {\scriptsize$\pm$0.2} \\
		&                                     & AnoFusion            & \cellcolor{yellow!20!green!35}1.8 {\scriptsize$\pm$0.0} & \cellcolor{yellow!14!green!35}23.5 {\scriptsize$\pm$0.4} & \cellcolor{yellow!31!green!35}7.0 {\scriptsize$\pm$0.2} \\
		&                                     & MSTGAD               & \cellcolor{red!100!yellow!35}5.8 {\scriptsize$\pm$0.1} & \cellcolor{red!100!yellow!35}89.5 {\scriptsize$\pm$0.8} & \cellcolor{red!100!yellow!35}21.6 {\scriptsize$\pm$0.8} \\
		&                                     & Art                  & \cellcolor{yellow!0!green!35}1.4 {\scriptsize$\pm$0.1} & \cellcolor{yellow!10!green!35}22.2 {\scriptsize$\pm$0.2} & \cellcolor{yellow!0!green!35}4.3 {\scriptsize$\pm$0.0} \\
		&                                     & Medicine             & \cellcolor{yellow!77!green!35}3.1 {\scriptsize$\pm$0.0} & \cellcolor{yellow!79!green!35}46.8 {\scriptsize$\pm$0.7} & \cellcolor{yellow!61!green!35}9.6 {\scriptsize$\pm$0.1} \\
		&                                     & OrEdge               & \cellcolor{yellow!16!green!35}1.7 {\scriptsize$\pm$0.0} & \cellcolor{yellow!11!green!35}22.8 {\scriptsize$\pm$0.4} & \cellcolor{yellow!29!green!35}6.9 {\scriptsize$\pm$0.1} \\
		\cmidrule{2-6}
		& FLOPs [M]                           & Eadro                & \cellcolor{yellow!17!green!35}185.8 {\scriptsize$\pm$0.000} & \cellcolor{yellow!9!green!35}459.3 {\scriptsize$\pm$0.000} & \cellcolor{yellow!36!green!35}50.3 {\scriptsize$\pm$0.000} \\
		&                                     & AnoFusion            & \cellcolor{yellow!32!green!35}310.4 {\scriptsize$\pm$0.000} & \cellcolor{yellow!40!green!35}1191.8 {\scriptsize$\pm$0.000} & \cellcolor{yellow!61!green!35}67.3 {\scriptsize$\pm$0.000} \\
		&                                     & MSTGAD               & \cellcolor{red!100!yellow!35}1725.7 {\scriptsize$\pm$0.000} & \cellcolor{red!100!yellow!35}4878.4 {\scriptsize$\pm$0.000} & \cellcolor{red!100!yellow!35}160.7 {\scriptsize$\pm$0.000} \\
		&                                     & Art                  & \cellcolor{yellow!5!green!35}82.1 {\scriptsize$\pm$0.000} & \cellcolor{yellow!7!green!35}417.7 {\scriptsize$\pm$0.000} & \cellcolor{yellow!0!green!35}26.0 {\scriptsize$\pm$0.000} \\
		&                                     & Medicine             & \cellcolor{yellow!27!green!35}270.3 {\scriptsize$\pm$0.000} & \cellcolor{yellow!18!green!35}681.5 {\scriptsize$\pm$0.000} & \cellcolor{yellow!28!green!35}45.0 {\scriptsize$\pm$0.000} \\
		&                                     & OrEdge               & \cellcolor{yellow!0!green!35}33.8 {\scriptsize$\pm$0.000} & \cellcolor{yellow!0!green!35}246.2 {\scriptsize$\pm$0.000} & \cellcolor{yellow!33!green!35}48.9 {\scriptsize$\pm$0.000} \\
		\cmidrule{2-6}
		& Reconstuction Params                & Eadro                & \cellcolor{red!100!yellow!35}110420.0 {\scriptsize$\pm$0.000} & \cellcolor{red!100!yellow!35}118953.0 {\scriptsize$\pm$0.000} & \cellcolor{red!100!yellow!35}143490.0 {\scriptsize$\pm$0.000} \\
		&                                     & AnoFusion            & \cellcolor{yellow!32!green!35}19897.0 {\scriptsize$\pm$0.000} & \cellcolor{yellow!47!green!35}32306.0 {\scriptsize$\pm$0.000} & \cellcolor{yellow!17!green!35}21603.0 {\scriptsize$\pm$0.000} \\
		&                                     & MSTGAD               & \cellcolor{red!67!yellow!35}92584.0 {\scriptsize$\pm$0.000} & \cellcolor{red!59!yellow!35}96029.0 {\scriptsize$\pm$0.000} & \cellcolor{yellow!40!green!35}36938.0 {\scriptsize$\pm$0.000} \\
		&                                     & Art                  & \cellcolor{yellow!35!green!35}21248.0 {\scriptsize$\pm$0.000} & \cellcolor{yellow!52!green!35}35231.0 {\scriptsize$\pm$0.000} & \cellcolor{yellow!24!green!35}25776.0 {\scriptsize$\pm$0.000} \\
		&                                     & Medicine             & \cellcolor{yellow!50!green!35}29458.0 {\scriptsize$\pm$0.000} & \cellcolor{yellow!51!green!35}34854.0 {\scriptsize$\pm$0.000} & \cellcolor{yellow!35!green!35}33477.0 {\scriptsize$\pm$0.000} \\
		&                                     & OrEdge               & \cellcolor{yellow!0!green!35}2179.0 {\scriptsize$\pm$0.000} & \cellcolor{yellow!0!green!35}5624.0 {\scriptsize$\pm$0.000} & \cellcolor{yellow!0!green!35}9581.0 {\scriptsize$\pm$0.000} \\
		\midrule[1.5pt]
		\smash{\begin{tabular}[t]{l}\textbf{Raspberry Pi 5}\\ \footnotesize (ARM Cortex-A76 16GB-Ram)\end{tabular}} & Inference Time (CPU) [ms]           & Eadro                & \cellcolor{yellow!2!green!35}78.28 {\scriptsize$\pm$0.52} & \cellcolor{yellow!1!green!35}171.39 {\scriptsize$\pm$3.08} & \cellcolor{yellow!0!green!35}22.89 {\scriptsize$\pm$0.10} \\
		&                                     & AnoFusion            & \cellcolor{yellow!57!green!35}869.60 {\scriptsize$\pm$15.41} & \cellcolor{yellow!35!green!35}1310.54 {\scriptsize$\pm$13.62} & \cellcolor{yellow!47!green!35}356.71 {\scriptsize$\pm$11.36} \\
		&                                     & MSTGAD               & \cellcolor{red!100!yellow!35}2907.40 {\scriptsize$\pm$5.98} & \cellcolor{red!100!yellow!35}6877.05 {\scriptsize$\pm$274.56} & \cellcolor{red!100!yellow!35}1432.36 {\scriptsize$\pm$16.84} \\
		&                                     & Art                  & \cellcolor{yellow!47!green!35}730.50 {\scriptsize$\pm$0.93} & \cellcolor{yellow!23!green!35}929.64 {\scriptsize$\pm$0.72} & \cellcolor{yellow!10!green!35}95.17 {\scriptsize$\pm$0.30} \\
		&                                     & Medicine             & \cellcolor{yellow!9!green!35}182.92 {\scriptsize$\pm$0.72} & \cellcolor{yellow!9!green!35}434.81 {\scriptsize$\pm$28.58} & \cellcolor{yellow!1!green!35}32.88 {\scriptsize$\pm$0.11} \\
		&                                     & OrEdge               & \cellcolor{yellow!0!green!35}45.63 {\scriptsize$\pm$0.21} & \cellcolor{yellow!0!green!35}125.42 {\scriptsize$\pm$2.98} & \cellcolor{yellow!1!green!35}30.67 {\scriptsize$\pm$1.27} \\
		\cmidrule{2-6}
		& Max Mem (CPU) [MB]                  & Eadro                & \cellcolor{yellow!46!green!35}530.3 {\scriptsize$\pm$1.7} & \cellcolor{yellow!26!green!35}654.7 {\scriptsize$\pm$3.6} & \cellcolor{yellow!84!green!35}870.6 {\scriptsize$\pm$3.2} \\
		&                                     & AnoFusion            & \cellcolor{yellow!21!green!35}500.3 {\scriptsize$\pm$0.5} & \cellcolor{yellow!11!green!35}597.5 {\scriptsize$\pm$7.4} & \cellcolor{yellow!3!green!35}860.4 {\scriptsize$\pm$0.1} \\
		&                                     & MSTGAD               & \cellcolor{red!100!yellow!35}720.0 {\scriptsize$\pm$25.9} & \cellcolor{red!100!yellow!35}1352.4 {\scriptsize$\pm$22.5} & \cellcolor{red!100!yellow!35}885.2 {\scriptsize$\pm$27.5} \\
		&                                     & Art                  & \cellcolor{yellow!21!green!35}500.3 {\scriptsize$\pm$1.2} & \cellcolor{yellow!11!green!35}597.5 {\scriptsize$\pm$6.5} & \cellcolor{yellow!0!green!35}860.0 {\scriptsize$\pm$2.4} \\
		&                                     & Medicine             & \cellcolor{yellow!7!green!35}482.9 {\scriptsize$\pm$0.5} & \cellcolor{yellow!0!green!35}549.7 {\scriptsize$\pm$5.5} & \cellcolor{yellow!5!green!35}860.7 {\scriptsize$\pm$2.3} \\
		&                                     & OrEdge               & \cellcolor{yellow!0!green!35}473.4 {\scriptsize$\pm$25.2} & \cellcolor{yellow!8!green!35}584.5 {\scriptsize$\pm$31.2} & \cellcolor{yellow!31!green!35}864.0 {\scriptsize$\pm$26.5} \\
		\midrule[1.5pt]
		\smash{\begin{tabular}[t]{l}\textbf{Raspberry Pi 3}\\ \footnotesize (ARM Cortex-A53 1GB-Ram)\end{tabular}} & Inference Time (CPU) [ms]           & Eadro                & \cellcolor{yellow!2!green!35}501.27 {\scriptsize$\pm$23.33} & \cellcolor{yellow!10!green!35}1147.67 {\scriptsize$\pm$25.46} & \cellcolor{yellow!0!green!35}145.00 {\scriptsize$\pm$0.49} \\
		&                                     & AnoFusion            & \cellcolor{yellow!50!green!35}4510.91 {\scriptsize$\pm$43.12} & \cellcolor{red!100!yellow!35}7281.65 {\scriptsize$\pm$75.58} & \cellcolor{yellow!44!green!35}1804.26 {\scriptsize$\pm$12.14} \\
		&                                     & MSTGAD               & \cellcolor{red!100!yellow!35}16973.47 {\scriptsize$\pm$346.74} & - & \cellcolor{red!100!yellow!35}7623.27 {\scriptsize$\pm$80.75} \\
		&                                     & Art                  & \cellcolor{yellow!50!green!35}4555.00 {\scriptsize$\pm$11.98} & \cellcolor{red!63!yellow!35}6085.58 {\scriptsize$\pm$13.16} & \cellcolor{yellow!11!green!35}583.29 {\scriptsize$\pm$0.40} \\
		&                                     & Medicine             & \cellcolor{yellow!10!green!35}1152.78 {\scriptsize$\pm$1.40} & \cellcolor{yellow!56!green!35}2617.04 {\scriptsize$\pm$7.58} & \cellcolor{yellow!1!green!35}212.29 {\scriptsize$\pm$0.79} \\
		&                                     & OrEdge               & \cellcolor{yellow!0!green!35}318.62 {\scriptsize$\pm$15.58} & \cellcolor{yellow!0!green!35}802.34 {\scriptsize$\pm$16.08} & \cellcolor{yellow!1!green!35}210.08 {\scriptsize$\pm$2.19} \\
		\cmidrule{2-6}
		& Max Mem (CPU) [MB]                  & Eadro                & \cellcolor{yellow!50!green!35}462.9 {\scriptsize$\pm$4.3} & \cellcolor{red!100!yellow!35}586.1 {\scriptsize$\pm$8.2} & \cellcolor{red!100!yellow!35}719.9 {\scriptsize$\pm$14.2} \\
		&                                     & AnoFusion            & \cellcolor{yellow!24!green!35}433.7 {\scriptsize$\pm$4.0} & \cellcolor{yellow!81!green!35}528.1 {\scriptsize$\pm$9.9} & \cellcolor{red!17!yellow!35}711.2 {\scriptsize$\pm$18.0} \\
		&                                     & MSTGAD               & \cellcolor{red!100!yellow!35}632.5 {\scriptsize$\pm$58.4} & - & \cellcolor{yellow!80!green!35}707.3 {\scriptsize$\pm$17.2} \\
		&                                     & Art                  & \cellcolor{yellow!26!green!35}435.4 {\scriptsize$\pm$5.4} & \cellcolor{yellow!36!green!35}506.5 {\scriptsize$\pm$34.6} & \cellcolor{yellow!19!green!35}700.8 {\scriptsize$\pm$20.2} \\
		&                                     & Medicine             & \cellcolor{yellow!0!green!35}406.6 {\scriptsize$\pm$16.1} & \cellcolor{yellow!0!green!35}488.5 {\scriptsize$\pm$5.3} & \cellcolor{yellow!0!green!35}698.8 {\scriptsize$\pm$15.2} \\
		&                                     & OrEdge               & \cellcolor{yellow!0!green!35}405.8 {\scriptsize$\pm$18.7} & \cellcolor{yellow!47!green!35}511.7 {\scriptsize$\pm$22.0} & \cellcolor{yellow!18!green!35}700.7 {\scriptsize$\pm$19.6} \\
		\bottomrule
	\end{tabular}
\end{table*}

Table~\ref{tab:efficiency_metrics} evaluates the computational efficiency and deployment feasibility of OrEdge compared with representative anomaly detection methods across server and edge devices. The evaluation considers training cost, FLOPs, reconstruction parameters, inference latency, and runtime memory usage.

\noindent\textbf{Model Complexity and Computational Cost.}
OrEdge introduces a lightweight temporal core based on orthogonal projection, orthogonal-domain filtering, and shared reconstruction modules, substantially reducing computational complexity compared with existing multi-modal approaches. 
On the server platform, OrEdge requires only 33.8M, 246.2M, and 48.9M FLOPs on SN, TT, and MSDS, respectively, achieving up to an order-of-magnitude reduction compared with graph-based methods such as MSTGAD (1725.7M--4878.4M FLOPs). 
Similarly, OrEdge reduces reconstruction parameters to 2.2k--9.6k, representing a $10\times$--$50\times$ reduction compared with multi-modal fusion approaches such as Eadro and MSTGAD. 
Despite this significant reduction in model complexity, OrEdge maintains comparable training time, requiring only 1.7s, 22.8s, and 6.9s across the three datasets, demonstrating that the proposed design achieves high computational efficiency without additional optimization overhead.

\noindent\textbf{Edge Deployment Efficiency.}
The lightweight architecture of OrEdge translates directly into improved inference efficiency on resource-constrained edge devices. 
On Raspberry Pi 5, OrEdge achieves inference latency of 30.67--125.42 ms per sample, outperforming all compared baselines. 
Compared with MSTGAD, OrEdge reduces inference latency by up to $64\times$, enabling real-time anomaly detection under highly constrained computing environments. 
Similar improvements are observed on the more constrained Raspberry Pi 3 platform, where OrEdge maintains inference latency below one second (210.08--802.34 ms). 
Notably, MSTGAD cannot be successfully deployed on Raspberry Pi 3 due to its substantially higher computational requirements, highlighting the practical advantage of OrEdge's lightweight design for constrained edge environments.

Runtime memory usage is measured using peak Resident Set Size (RSS) during inference. While RSS captures the complete deployment process, including framework and runtime overhead, OrEdge maintains a stable memory footprint across different devices (473--864 MB on Raspberry Pi 5 and 406--701 MB on Raspberry Pi 3), demonstrating its feasibility for deployment on resource-constrained platforms. 
These results demonstrate that the proposed orthogonal projection-based temporal core fundamentally changes the accuracy-efficiency trade-off. Instead of relying on computationally expensive Transformer attention or graph neural network message passing, OrEdge exploits orthogonal representations to enable lightweight linear temporal modeling while preserving the temporal and cross-modal information required for anomaly detection.

\begin{tcolorbox}[
	colback=gray!10,
	colframe=black!50,
	title={RQ2 Summary: Efficiency and Edge Deployment Analysis},
	left=0mm,
	right=1mm,
	top=1mm,
	bottom=1mm,
	boxsep=1mm,
	arc=2mm
	]
	\begin{itemize}
		\item OrEdge significantly reduces computational complexity and resource consumption compared with existing multi-modal anomaly detection methods,
		 enabling efficient anomaly detection with minimal resource overhead.
		
		\item OrEdge allows practical deployment on resource-constrained Raspberry Pi platforms.
	\end{itemize}
\end{tcolorbox}


\subsubsection{\textbf{(RQ3) Ablation Study Analysis}}

\begin{table*}[t]
	\centering
	\caption{Ablation Analysis of Architectural Components and Orthogonal Transformation Bases (Mean $\pm$ SD)}
	\label{tab:comprehensive_rq2}
	\small
	\setlength{\tabcolsep}{3pt}
	\begin{tabular}{l|ccccc | ccccc | ccccc}
		\toprule
		\textbf{Configuration Group} & \multicolumn{5}{c}{\textbf{TT}} & \multicolumn{5}{c}{\textbf{SN}} & \multicolumn{5}{c}{\textbf{MSDS*}} \\
		\cmidrule(lr){2-6} \cmidrule(lr){7-11} \cmidrule(lr){12-16}
		& PR & RC & F1 & AUC & AP & PR & RC & F1 & AUC & AP & PR & RC & F1 & AUC & AP \\
		\midrule
		\multicolumn{16}{l}{\textbf{(a) Impact of Architectural Components}} \\
		without FreDF loss                                 & \cellcolor{green!53!yellow!35}\shortstack{0.928\\ \tiny$\pm$0.005} & \cellcolor{yellow!0!red!35}\shortstack{0.797\\ \tiny$\pm$0.029} & \cellcolor{yellow!0!red!35}\shortstack{0.857\\ \tiny$\pm$0.016} & \cellcolor{yellow!0!red!35}\shortstack{0.997\\ \tiny$\pm$0.000} & \cellcolor{yellow!0!red!35}\shortstack{0.968\\ \tiny$\pm$0.004} & \cellcolor{yellow!0!red!35}\shortstack{0.945\\ \tiny$\pm$0.001} & \cellcolor{yellow!0!red!35}\shortstack{0.919\\ \tiny$\pm$0.012} & \cellcolor{yellow!0!red!35}\shortstack{0.932\\ \tiny$\pm$0.006} & \cellcolor{yellow!66!red!35}\shortstack{0.994\\ \tiny$\pm$0.002} & \cellcolor{yellow!0!red!35}\shortstack{0.984\\ \tiny$\pm$0.002} & -- & -- & -- & -- & -- \\
		without linear-attn                                & \cellcolor{yellow!0!red!35}\shortstack{0.874\\ \tiny$\pm$0.035} & \cellcolor{yellow!87!red!35}\shortstack{0.874\\ \tiny$\pm$0.018} & \cellcolor{yellow!33!red!35}\shortstack{0.874\\ \tiny$\pm$0.009} & \cellcolor{yellow!23!red!35}\shortstack{0.997\\ \tiny$\pm$0.000} & \cellcolor{yellow!39!red!35}\shortstack{0.973\\ \tiny$\pm$0.004} & \cellcolor{yellow!75!red!35}\shortstack{0.955\\ \tiny$\pm$0.005} & \cellcolor{yellow!70!red!35}\shortstack{0.933\\ \tiny$\pm$0.014} & \cellcolor{yellow!72!red!35}\shortstack{0.944\\ \tiny$\pm$0.007} & \cellcolor{yellow!0!red!35}\shortstack{0.992\\ \tiny$\pm$0.002} & \cellcolor{yellow!26!red!35}\shortstack{0.985\\ \tiny$\pm$0.003} & \cellcolor{yellow!0!red!35}\shortstack{0.919\\ \tiny$\pm$0.055} & \cellcolor{yellow!0!red!35}\shortstack{0.974\\ \tiny$\pm$0.009} & \cellcolor{yellow!0!red!35}\shortstack{0.946\\ \tiny$\pm$0.032} & \cellcolor{green!92!yellow!35}\shortstack{0.998\\ \tiny$\pm$0.002} & \cellcolor{yellow!65!red!35}\shortstack{0.974\\ \tiny$\pm$0.009} \\
		without NormLin                                    & \cellcolor{green!85!yellow!35}\shortstack{0.939\\ \tiny$\pm$0.005} & \cellcolor{green!100!yellow!35}\shortstack{0.973\\ \tiny$\pm$0.002} & \cellcolor{green!100!yellow!35}\shortstack{0.956\\ \tiny$\pm$0.003} & \cellcolor{green!100!yellow!35}\shortstack{1.000\\ \tiny$\pm$0.000} & \cellcolor{green!100!yellow!35}\shortstack{0.995\\ \tiny$\pm$0.000} & \cellcolor{green!91!yellow!35}\shortstack{0.970\\ \tiny$\pm$0.000} & \cellcolor{green!78!yellow!35}\shortstack{0.954\\ \tiny$\pm$0.000} & \cellcolor{green!83!yellow!35}\shortstack{0.962\\ \tiny$\pm$0.000} & \cellcolor{green!100!yellow!35}\shortstack{0.996\\ \tiny$\pm$0.000} & \cellcolor{green!100!yellow!35}\shortstack{0.992\\ \tiny$\pm$0.000} & \cellcolor{green!85!yellow!35}\shortstack{0.948\\ \tiny$\pm$0.000} & \cellcolor{green!100!yellow!35}\shortstack{0.985\\ \tiny$\pm$0.000} & \cellcolor{green!100!yellow!35}\shortstack{0.966\\ \tiny$\pm$0.000} & \cellcolor{green!100!yellow!35}\shortstack{0.998\\ \tiny$\pm$0.000} & \cellcolor{yellow!0!red!35}\shortstack{0.973\\ \tiny$\pm$0.000} \\
		OrEdge (linear attn, norm, FreDF)                  & \cellcolor{green!100!yellow!35}\shortstack{0.944\\ \tiny$\pm$0.002} & \cellcolor{green!85!yellow!35}\shortstack{0.961\\ \tiny$\pm$0.007} & \cellcolor{green!93!yellow!35}\shortstack{0.952\\ \tiny$\pm$0.003} & \cellcolor{green!100!yellow!35}\shortstack{1.000\\ \tiny$\pm$0.000} & \cellcolor{green!95!yellow!35}\shortstack{0.994\\ \tiny$\pm$0.001} & \cellcolor{green!100!yellow!35}\shortstack{0.971\\ \tiny$\pm$0.005} & \cellcolor{green!100!yellow!35}\shortstack{0.958\\ \tiny$\pm$0.008} & \cellcolor{green!100!yellow!35}\shortstack{0.965\\ \tiny$\pm$0.004} & \cellcolor{yellow!96!red!35}\shortstack{0.994\\ \tiny$\pm$0.001} & \cellcolor{green!89!yellow!35}\shortstack{0.992\\ \tiny$\pm$0.002} & \cellcolor{green!100!yellow!35}\shortstack{0.950\\ \tiny$\pm$0.007} & \cellcolor{green!49!yellow!35}\shortstack{0.982\\ \tiny$\pm$0.010} & \cellcolor{green!98!yellow!35}\shortstack{0.966\\ \tiny$\pm$0.003} & \cellcolor{yellow!0!red!35}\shortstack{0.994\\ \tiny$\pm$0.006} & \cellcolor{green!100!yellow!35}\shortstack{0.978\\ \tiny$\pm$0.005} \\
		\midrule[2pt]
		\multicolumn{16}{l}{\textbf{(b) Impact of Alternative Functional Projection Bases}} \\
		Hermite Basis                                      & \cellcolor{yellow!0!red!35}\shortstack{0.698\\ \tiny$\pm$0.049} & \cellcolor{yellow!0!red!35}\shortstack{0.720\\ \tiny$\pm$0.054} & \cellcolor{yellow!0!red!35}\shortstack{0.707\\ \tiny$\pm$0.022} & \cellcolor{yellow!0!red!35}\shortstack{0.893\\ \tiny$\pm$0.014} & \cellcolor{yellow!0!red!35}\shortstack{0.799\\ \tiny$\pm$0.013} & \cellcolor{yellow!0!red!35}\shortstack{0.352\\ \tiny$\pm$0.041} & \cellcolor{yellow!0!red!35}\shortstack{0.544\\ \tiny$\pm$0.063} & \cellcolor{yellow!0!red!35}\shortstack{0.428\\ \tiny$\pm$0.050} & \cellcolor{yellow!0!red!35}\shortstack{0.754\\ \tiny$\pm$0.033} & \cellcolor{yellow!0!red!35}\shortstack{0.605\\ \tiny$\pm$0.023} & \cellcolor{yellow!1!red!35}\shortstack{0.890\\ \tiny$\pm$0.043} & \cellcolor{yellow!0!red!35}\shortstack{0.920\\ \tiny$\pm$0.016} & \cellcolor{yellow!0!red!35}\shortstack{0.904\\ \tiny$\pm$0.015} & \cellcolor{yellow!0!red!35}\shortstack{0.989\\ \tiny$\pm$0.008} & \cellcolor{yellow!3!red!35}\shortstack{0.959\\ \tiny$\pm$0.009} \\
		Chebyshev Basis                                    & \cellcolor{green!87!yellow!35}\shortstack{0.928\\ \tiny$\pm$0.010} & \cellcolor{green!96!yellow!35}\shortstack{0.957\\ \tiny$\pm$0.020} & \cellcolor{green!91!yellow!35}\shortstack{0.942\\ \tiny$\pm$0.013} & \cellcolor{green!99!yellow!35}\shortstack{0.999\\ \tiny$\pm$0.000} & \cellcolor{green!98!yellow!35}\shortstack{0.992\\ \tiny$\pm$0.002} & \cellcolor{green!95!yellow!35}\shortstack{0.957\\ \tiny$\pm$0.002} & \cellcolor{green!95!yellow!35}\shortstack{0.949\\ \tiny$\pm$0.011} & \cellcolor{green!95!yellow!35}\shortstack{0.953\\ \tiny$\pm$0.006} & \cellcolor{green!98!yellow!35}\shortstack{0.994\\ \tiny$\pm$0.000} & \cellcolor{green!98!yellow!35}\shortstack{0.989\\ \tiny$\pm$0.002} & \cellcolor{green!45!yellow!35}\shortstack{0.934\\ \tiny$\pm$0.006} & \cellcolor{green!53!yellow!35}\shortstack{0.979\\ \tiny$\pm$0.009} & \cellcolor{green!68!yellow!35}\shortstack{0.956\\ \tiny$\pm$0.002} & \cellcolor{green!75!yellow!35}\shortstack{0.999\\ \tiny$\pm$0.001} & \cellcolor{green!78!yellow!35}\shortstack{0.980\\ \tiny$\pm$0.006} \\
		Laguerre Basis                                     & \cellcolor{green!85!yellow!35}\shortstack{0.926\\ \tiny$\pm$0.030} & \cellcolor{green!83!yellow!35}\shortstack{0.941\\ \tiny$\pm$0.033} & \cellcolor{green!84!yellow!35}\shortstack{0.933\\ \tiny$\pm$0.029} & \cellcolor{green!99!yellow!35}\shortstack{0.999\\ \tiny$\pm$0.001} & \cellcolor{green!96!yellow!35}\shortstack{0.990\\ \tiny$\pm$0.007} & \cellcolor{green!97!yellow!35}\shortstack{0.964\\ \tiny$\pm$0.004} & \cellcolor{green!92!yellow!35}\shortstack{0.942\\ \tiny$\pm$0.020} & \cellcolor{green!95!yellow!35}\shortstack{0.953\\ \tiny$\pm$0.013} & \cellcolor{green!100!yellow!35}\shortstack{0.995\\ \tiny$\pm$0.001} & \cellcolor{green!100!yellow!35}\shortstack{0.992\\ \tiny$\pm$0.002} & \cellcolor{yellow!0!red!35}\shortstack{0.889\\ \tiny$\pm$0.034} & \cellcolor{yellow!13!red!35}\shortstack{0.925\\ \tiny$\pm$0.012} & \cellcolor{yellow!8!red!35}\shortstack{0.907\\ \tiny$\pm$0.019} & \cellcolor{yellow!3!red!35}\shortstack{0.989\\ \tiny$\pm$0.001} & \cellcolor{yellow!0!red!35}\shortstack{0.958\\ \tiny$\pm$0.006} \\
		Fourier Basis                                      & \cellcolor{green!86!yellow!35}\shortstack{0.927\\ \tiny$\pm$0.008} & \cellcolor{yellow!44!red!35}\shortstack{0.774\\ \tiny$\pm$0.022} & \cellcolor{green!11!yellow!35}\shortstack{0.844\\ \tiny$\pm$0.016} & \cellcolor{green!93!yellow!35}\shortstack{0.996\\ \tiny$\pm$0.001} & \cellcolor{green!67!yellow!35}\shortstack{0.962\\ \tiny$\pm$0.006} & \cellcolor{green!93!yellow!35}\shortstack{0.952\\ \tiny$\pm$0.005} & \cellcolor{green!73!yellow!35}\shortstack{0.903\\ \tiny$\pm$0.009} & \cellcolor{green!85!yellow!35}\shortstack{0.927\\ \tiny$\pm$0.003} & \cellcolor{green!96!yellow!35}\shortstack{0.992\\ \tiny$\pm$0.001} & \cellcolor{green!95!yellow!35}\shortstack{0.983\\ \tiny$\pm$0.001} & \cellcolor{green!20!yellow!35}\shortstack{0.926\\ \tiny$\pm$0.014} & \cellcolor{green!100!yellow!35}\shortstack{0.997\\ \tiny$\pm$0.005} & \cellcolor{green!82!yellow!35}\shortstack{0.960\\ \tiny$\pm$0.005} & \cellcolor{green!100!yellow!35}\shortstack{1.000\\ \tiny$\pm$0.000} & \cellcolor{green!100!yellow!35}\shortstack{0.983\\ \tiny$\pm$0.005} \\
		OrEdge (with legendre)                             & \cellcolor{green!100!yellow!35}\shortstack{0.944\\ \tiny$\pm$0.002} & \cellcolor{green!100!yellow!35}\shortstack{0.961\\ \tiny$\pm$0.007} & \cellcolor{green!100!yellow!35}\shortstack{0.952\\ \tiny$\pm$0.003} & \cellcolor{green!100!yellow!35}\shortstack{1.000\\ \tiny$\pm$0.000} & \cellcolor{green!100!yellow!35}\shortstack{0.994\\ \tiny$\pm$0.001} & \cellcolor{green!100!yellow!35}\shortstack{0.971\\ \tiny$\pm$0.005} & \cellcolor{green!100!yellow!35}\shortstack{0.958\\ \tiny$\pm$0.008} & \cellcolor{green!100!yellow!35}\shortstack{0.965\\ \tiny$\pm$0.004} & \cellcolor{green!99!yellow!35}\shortstack{0.994\\ \tiny$\pm$0.001} & \cellcolor{green!99!yellow!35}\shortstack{0.992\\ \tiny$\pm$0.002} & \cellcolor{green!100!yellow!35}\shortstack{0.950\\ \tiny$\pm$0.008} & \cellcolor{green!60!yellow!35}\shortstack{0.982\\ \tiny$\pm$0.012} & \cellcolor{green!100!yellow!35}\shortstack{0.966\\ \tiny$\pm$0.004} & \cellcolor{green!0!yellow!35}\shortstack{0.994\\ \tiny$\pm$0.006} & \cellcolor{green!60!yellow!35}\shortstack{0.978\\ \tiny$\pm$0.006} \\
		\midrule[2pt]
		\multicolumn{16}{l}{\textbf{(c) Impact of Architectural Variants}} \\
		FEDformerModel                                     & \cellcolor{yellow!0!red!35}\shortstack{0.793\\ \tiny$\pm$0.013} & \cellcolor{yellow!0!red!35}\shortstack{0.590\\ \tiny$\pm$0.009} & \cellcolor{yellow!0!red!35}\shortstack{0.676\\ \tiny$\pm$0.011} & \cellcolor{yellow!0!red!35}\shortstack{0.984\\ \tiny$\pm$0.001} & \cellcolor{yellow!0!red!35}\shortstack{0.892\\ \tiny$\pm$0.005} & \cellcolor{yellow!0!red!35}\shortstack{0.840\\ \tiny$\pm$0.023} & \cellcolor{yellow!0!red!35}\shortstack{0.774\\ \tiny$\pm$0.018} & \cellcolor{yellow!0!red!35}\shortstack{0.806\\ \tiny$\pm$0.014} & \cellcolor{yellow!0!red!35}\shortstack{0.981\\ \tiny$\pm$0.002} & \cellcolor{yellow!0!red!35}\shortstack{0.940\\ \tiny$\pm$0.006} & \cellcolor{yellow!0!red!35}\shortstack{0.892\\ \tiny$\pm$0.010} & \cellcolor{yellow!0!red!35}\shortstack{0.767\\ \tiny$\pm$0.011} & \cellcolor{yellow!0!red!35}\shortstack{0.825\\ \tiny$\pm$0.002} & \cellcolor{yellow!0!red!35}\shortstack{0.962\\ \tiny$\pm$0.023} & \cellcolor{yellow!0!red!35}\shortstack{0.906\\ \tiny$\pm$0.014} \\
		iTransformer                                       & \cellcolor{green!83!yellow!35}\shortstack{0.931\\ \tiny$\pm$0.009} & \cellcolor{green!98!yellow!35}\shortstack{0.959\\ \tiny$\pm$0.003} & \cellcolor{green!94!yellow!35}\shortstack{0.945\\ \tiny$\pm$0.006} & \cellcolor{green!97!yellow!35}\shortstack{0.999\\ \tiny$\pm$0.000} & \cellcolor{green!97!yellow!35}\shortstack{0.993\\ \tiny$\pm$0.001} & \cellcolor{green!46!yellow!35}\shortstack{0.936\\ \tiny$\pm$0.020} & \cellcolor{green!59!yellow!35}\shortstack{0.921\\ \tiny$\pm$0.030} & \cellcolor{green!54!yellow!35}\shortstack{0.929\\ \tiny$\pm$0.024} & \cellcolor{green!46!yellow!35}\shortstack{0.991\\ \tiny$\pm$0.003} & \cellcolor{green!54!yellow!35}\shortstack{0.980\\ \tiny$\pm$0.006} & \cellcolor{green!9!yellow!35}\shortstack{0.924\\ \tiny$\pm$0.004} & \cellcolor{green!15!yellow!35}\shortstack{0.891\\ \tiny$\pm$0.011} & \cellcolor{green!17!yellow!35}\shortstack{0.907\\ \tiny$\pm$0.004} & \cellcolor{green!66!yellow!35}\shortstack{0.989\\ \tiny$\pm$0.007} & \cellcolor{green!81!yellow!35}\shortstack{0.972\\ \tiny$\pm$0.005} \\
		OrEdge                                             & \cellcolor{green!100!yellow!35}\shortstack{0.944\\ \tiny$\pm$0.002} & \cellcolor{green!100!yellow!35}\shortstack{0.961\\ \tiny$\pm$0.007} & \cellcolor{green!100!yellow!35}\shortstack{0.952\\ \tiny$\pm$0.003} & \cellcolor{green!100!yellow!35}\shortstack{1.000\\ \tiny$\pm$0.000} & \cellcolor{green!100!yellow!35}\shortstack{0.994\\ \tiny$\pm$0.001} & \cellcolor{green!100!yellow!35}\shortstack{0.971\\ \tiny$\pm$0.005} & \cellcolor{green!100!yellow!35}\shortstack{0.958\\ \tiny$\pm$0.008} & \cellcolor{green!100!yellow!35}\shortstack{0.965\\ \tiny$\pm$0.004} & \cellcolor{green!100!yellow!35}\shortstack{0.994\\ \tiny$\pm$0.001} & \cellcolor{green!100!yellow!35}\shortstack{0.992\\ \tiny$\pm$0.002} & \cellcolor{green!100!yellow!35}\shortstack{0.950\\ \tiny$\pm$0.008} & \cellcolor{green!100!yellow!35}\shortstack{0.982\\ \tiny$\pm$0.012} & \cellcolor{green!100!yellow!35}\shortstack{0.966\\ \tiny$\pm$0.004} & \cellcolor{green!100!yellow!35}\shortstack{0.994\\ \tiny$\pm$0.006} & \cellcolor{green!100!yellow!35}\shortstack{0.978\\ \tiny$\pm$0.006} \\
		\bottomrule
	\end{tabular}
	\begin{flushleft}
		*For MSDS, from RQ4, we find that orthogonal-domain supervision provides limited benefit, so we omit the FreDF loss in the ablation study, all experiments are conducted with $\lambda_{aux}=0$. 
	\end{flushleft}
\end{table*}
Table~\ref{tab:comprehensive_rq2} evaluates the contribution of individual OrEdge components, orthogonal projection choices, and alternative temporal modeling architectures. 

\noindent\textbf{Impact of Individual Components.}
The results demonstrate that each component contributes to OrEdge performance. Removing the orthogonal-domain objective degrades detection accuracy, particularly on datasets with richer temporal dependencies, confirming the benefit of orthogonal-aware optimization. Similarly, linear attention and NormLin improve temporal dependency modeling and feature stability, respectively, highlighting their complementary roles in multi-modal anomaly detection.

\noindent\textbf{Effect of Orthogonal Projection Bases.}
We further evaluate different orthogonal projection spaces by replacing the Legendre basis with Hermite, Chebyshev, Laguerre, and Fourier representations. The results show that the choice of basis significantly influences detection performance, with Legendre providing the most stable and effective representation across datasets. This indicates that suitable orthogonal projections can preserve important temporal structures while maintaining compact representations.

\noindent\textbf{Comparison with Alternative Temporal Architectures.}
Finally, we compare OrEdge with representative Transformer-based temporal models. While attention-based architectures achieve competitive performance, they introduce substantially higher computational complexity. In contrast, OrEdge achieves accurate anomaly detection using lightweight temporal operators, demonstrating that orthogonal-domain representations can provide an efficient alternative to complex sequence models for resource-constrained deployments.
\begin{tcolorbox}[
	colback=gray!10,
	colframe=black!50,
	title={RQ3 Summary: Ablation Study},
	left=0mm,
	right=1mm,
	top=1mm,
	bottom=1mm,
	boxsep=1mm,
	arc=2mm
	]
	\small
	\begin{itemize}
		\item Ablation results confirm that orthogonal-domain modeling and its components provide the best balance between anomaly detection accuracy and computational efficiency.
		\item The choice of orthogonal projection basis significantly affects performance, with Legendre providing the most stable and effective representation across datasets.
	\end{itemize}
\end{tcolorbox}

\subsubsection{\textbf{(RQ4) Sensitivity Analysis}}
To evaluate the robustness of OrEdge under different design choices, we analyze the sensitivity of two key parameters: the latent dimension of the linear attention module and the weight of the auxiliary orthogonal-domain reconstruction loss. Experiments are conducted under identical settings across all datasets, with final configurations selected based on the balance between detection performance and computational efficiency.

\begin{figure*}[t]
	\centering
	
	\subfloat[SN Dataset\label{fig:sens_linear_attn_sn}]{%
		\includegraphics[width=0.25\textwidth]{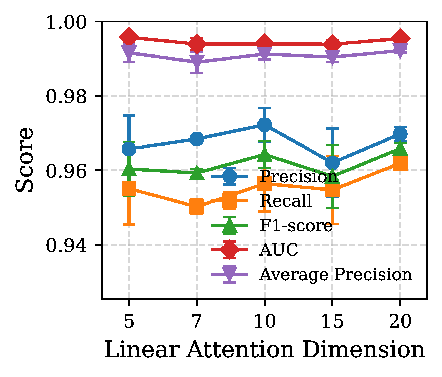}
	}
	\hfill
	\subfloat[MSDS Dataset\label{fig:sens_linear_attn_msds}]{%
		\includegraphics[width=0.25\textwidth]{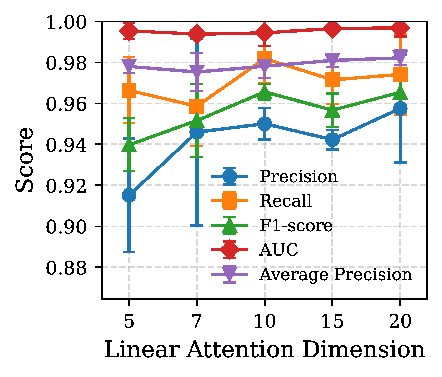}
	}
	\hfill
	\subfloat[TT Dataset\label{fig:sens_linear_attn_tt}]{%
		\includegraphics[width=0.25\textwidth]{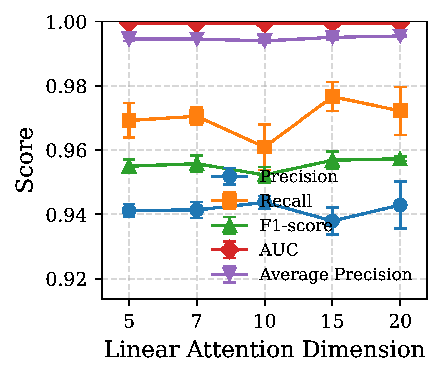}
	}
	
	\caption{Sensitivity analysis of the linear attention dimension. Each subplot reports the mean Precision (PR), Recall (RC), F1-score, AUC and average precision (AP) under different linear attention dimensions, with error bars indicating  standard deviation over repeated runs.}
	\label{fig:parameter_sensitivity_linear_attn_dim}
\end{figure*}

\noindent\textbf{Effect of Linear Attention Dimension.}
We vary the latent dimension of the linear attention module across $\{5,7,10,15,20\}$ to evaluate its impact on temporal representation capacity. As shown in Fig.~\ref{fig:parameter_sensitivity_linear_attn_dim}, OrEdge maintains stable performance across different dimensions, demonstrating that the proposed orthogonal temporal representation can effectively capture temporal dependencies using a compact latent space. Therefore, a dimension of 10 is selected as it provides a favorable trade-off between accuracy and efficiency.

\noindent\textbf{Effect of Orthogonal-Domain Reconstruction Weight.}
We further investigate the impact of the auxiliary orthogonal-domain reconstruction loss by varying $\lambda_{aux}$. The results show that its contribution depends on the temporal characteristics of each workload. Datasets with richer temporal patterns (i.e., TT and SN) benefit from stronger orthogonal-domain supervision, whereas simpler workloads such as MSDS show limited improvement. These findings demonstrate that orthogonal-aware optimization improves representation quality while avoiding unnecessary computational constraints when additional supervision provides limited benefit.

\begin{tcolorbox}[
	colback=gray!10,
	colframe=black!50,
	title={RQ4 Summary: Sensitivity Analysis},
	left=0mm,
	right=1mm,
	top=1mm,
	bottom=1mm,
	boxsep=1mm,
	arc=2mm
	]
	\small
	\begin{itemize}
		\item A compact linear attention dimension is sufficient for capturing temporal dependencies, indicating that the orthogonal temporal representation is effective even with limited latent capacity.
		\item OrEdge remains robust across different temporal capacities, while orthogonal optimization provides adaptive benefits depending on dataset temporal characteristics.
	\end{itemize}
\end{tcolorbox}

\subsubsection{\textbf{(RQ5) Case Study}}

\begin{figure*}[t]
	\centering
	
	\subfloat[SN Dataset\label{fig:sens_auxi_lambda_sn}]{%
		\includegraphics[width=0.25\textwidth]{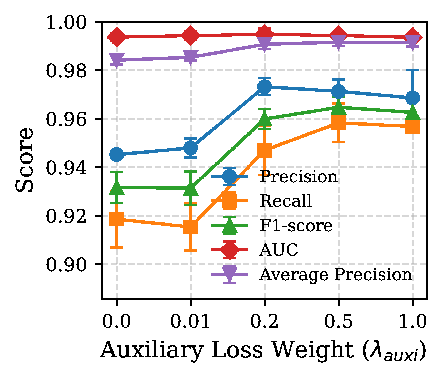}
	}
	\hfill
	\subfloat[MSDS Dataset\label{fig:sens_auxi_lambda_msds}]{%
		\includegraphics[width=0.25\textwidth]{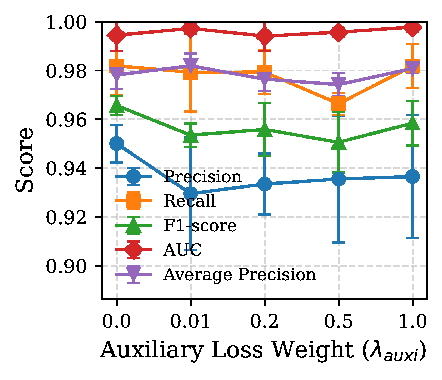}
	}
	\hfill
	\subfloat[TT Dataset\label{fig:sens_auxi_lambda_tt}]{%
		\includegraphics[width=0.25\textwidth]{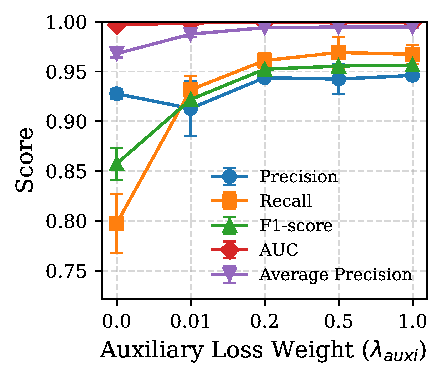}
	}

	\caption{Sensitivity analysis of the auxiliary loss weight ($\lambda_{auxi}$) across the SN, MSDS, and TT datasets. Each subplot reports the mean Precision (PR), Recall (RC), F1-score, AUC and average precision (AP) under different auxiliary loss weights, with error bars indicating  standard deviation over repeated runs.}
	\label{fig:parameter_sensitivity_auxi}
\end{figure*}

\begin{figure*}[t]
	\centering
	
	\begin{tcolorbox}[
		colback=gray!5,
		colframe=gray!40,
		boxrule=0.4pt,
		arc=2mm,
		title=\textbf{Eadro vs. OrEdge},
		coltitle=black,
		coltext=black,
		fonttitle=\bfseries
		]
		
		\begin{minipage}[t]{0.48\textwidth}
			\centering
			\includegraphics[width=\linewidth]{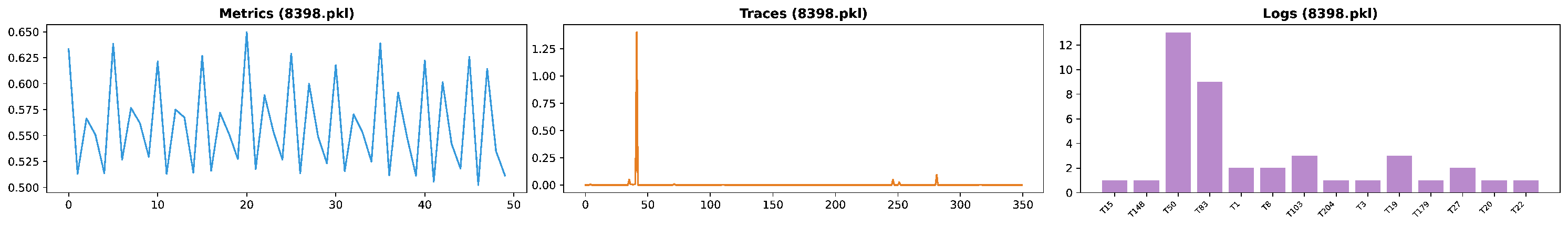}
			
			\vspace{1mm}
			\small (a) Detected by both methods.
			\label{fig:case-a}
		\end{minipage}
		\hfill
		\begin{minipage}[t]{0.48\textwidth}
			\centering
			\includegraphics[width=\linewidth]{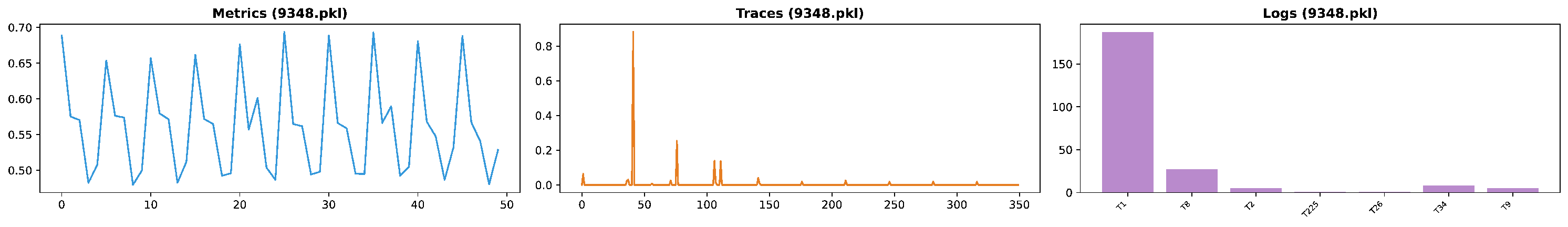}
			
			\vspace{1mm}
			\small (b) Correctly detected only by OrEdge.
			\label{fig:case-b}
		\end{minipage}
		
	\end{tcolorbox}

	\vspace{4mm}

	\begin{tcolorbox}[
		colback=gray!5,
		colframe=gray!40,
		boxrule=0.4pt,
		arc=2mm,
		title=\textbf{AnoFusion vs. OrEdge},
		coltitle=black,
		coltext=black,
		fonttitle=\bfseries
		]
		
		\begin{minipage}[t]{0.48\textwidth}
			\centering
			\includegraphics[width=\linewidth]{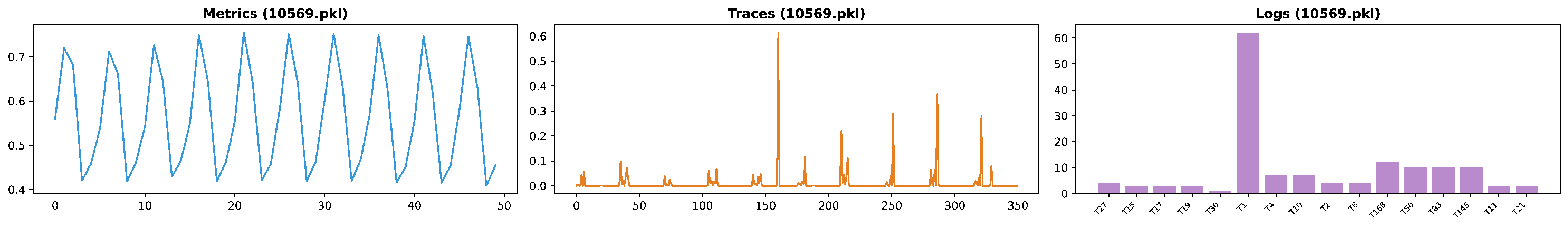}
			
			\vspace{1mm}
			\small (c) Detected by both methods.
			\label{fig:case-c}
		\end{minipage}
		\hfill
		\begin{minipage}[t]{0.48\textwidth}
			\centering
			\includegraphics[width=\linewidth]{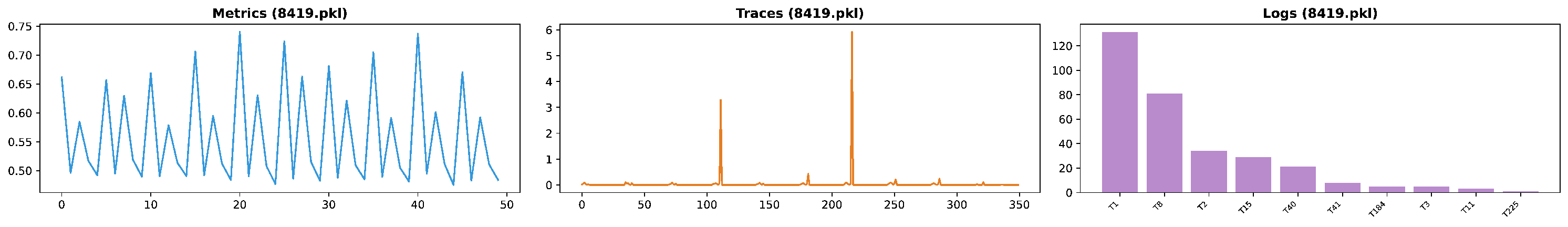}
			
			\vspace{1mm}
			\small (d) Correctly detected only by OrEdge.
			\label{fig:case-d}
		\end{minipage}
		
	\end{tcolorbox}
	
	\caption{
		Qualitative comparison of OrEdge against Eadro and AnoFusion.
		Each comparison presents cases detected by both methods and cases where
		OrEdge successfully identifies anomalies missed by the baselines.
	}
	\label{fig:casestudy}
\end{figure*}

We qualitatively compare OrEdge against the state-of-the-art Eadro and AnoFusion baselines on the MSDS dataset using the same training protocol as in RQ1. Consistent with the quantitative results in Table~\ref{tab:accuracy_metrics}, OrEdge achieves the highest F1-score (0.966) while requiring only 9.6K parameters, compared with 143K for Eadro and 21K for AnoFusion, highlighting its strong accuracy--efficiency trade-off.
To better understand the behavioral differences between the models, we examine representative windows where their predictions differ. For each baseline, we present two cases: (i) a window correctly identified by both methods, and (ii) a challenging window correctly detected only by OrEdge (Fig.~\ref{fig:casestudy}). Metrics are visualized over 50 temporal samples, traces correspond to the flattened node-to-node interaction matrix across the observation window, and logs are represented by template frequencies. Representative cases are selected according to the highest anomaly scores produced by each model.
Across both baselines, windows correctly identified by all methods exhibit pronounced metric deviations accompanied by widespread log-template activations, indicating clear system-wide failures. In contrast, windows missed by the baselines but detected by OrEdge contain weaker and more localized signatures: metric variations are less pronounced, log activations are sparse, and trace patterns remain visually similar. For instance, in case (b), the metric signals exhibit only subtle deviations between 20 and 40 seconds, while the corresponding log templates contain sparse and intermittent activations. Similarly, in case (d), the metric values initially remain within a narrow range around 0.55 before gradually increasing to approximately 0.75, producing a broader but still relatively smooth variation compared with the more distinctive patterns observed in the easier cases, while the log signals exhibit only limited activity. These weak and localized patterns are challenging to identify using representations that primarily rely on prominent time-domain variations. In contrast, OrEdge successfully detects both anomalies by leveraging its Legendre-domain representation, which captures underlying temporal structures and distributional deviations that are less evident in the original time domain. These observations suggest that OrEdge's Legendre-domain representation is more sensitive to subtle structural and temporal anomalies, enabling reliable detection even when conventional time-domain patterns are ambiguous.

\begin{tcolorbox}[
    colback=gray!10,
    colframe=black!50,
    title={RQ5 Summary: Case Study},
    left=0mm,        
    right=1mm,       
    top=1mm,         
    bottom=1mm,      
    boxsep=1mm,      
    arc=2mm          
]    \begin{itemize}
		\item OrEdge demonstrates superior anomaly detection capabilities compared to Eadro and AnoFusion, particularly in cases with subtle and distributed failure patterns, while maintaining extreme parameter efficiency.
    \end{itemize}
\end{tcolorbox}

	\section{Conclusion}

This study addresses the challenge of achieving accurate multi-modal anomaly detection under strict computational and deployment constraints. Existing approaches often rely on graph-based and attention-driven architectures that provide strong detection capability but introduce substantial computational and memory overhead, limiting their applicability in real-time edge environments.
To address this challenge, we proposed OrEdge, a lightweight orthogonal-domain framework that jointly analyzes metrics, logs, and traces through compact temporal representations. By leveraging orthogonal projections and efficient temporal modeling, OrEdge captures cross-modal anomalies while avoiding the high complexity of conventional graph and attention-based approaches.
Extensive experiments on three real-world microservice datasets demonstrate that OrEdge achieves competitive detection performance while reducing reconstruction parameters by up to an order of magnitude compared with existing multi-modal approaches. Furthermore, OrEdge enables practical deployment on resource-constrained Raspberry Pi platforms, achieving sub-second inference and substantial latency reductions while maintaining comparable detection accuracy.
Ablation studies, sensitivity analysis, and orthogonal basis evaluations validate the contribution of each component, while qualitative case studies demonstrate OrEdge's ability to identify subtle distributed anomalies. Future work will investigate the integration of advanced state-space models, such as Mamba, to further enhance long-range dependency modeling and scalability in distributed edge environments.

	\bibliographystyle{IEEEtran}
	\bibliography{bibfile}
	
	
\end{document}